\documentclass[acmsmall]{acmart}

\usepackage{todonotes}
\usepackage{soul}
\usepackage{amsmath}
\usepackage{wrapfig}

\theoremstyle{definition}
\newtheorem{example}{Example}[section]

\usepackage{xcolor} 
\newcommand{\newtext}[1]{\textcolor{black}{#1}}

\graphicspath{{figures/}}

\begin{document}
\title{Methodology of Algorithm Engineering}
\author{Jan Mendling}
\authornote{Corresponding Author}
\email{jan.mendling@hu-berlin.de}
\orcid{0000-0002-7260-524X}
\affiliation{%
  \institution{Humboldt-Universit{\"a}t zu Berlin}
  \streetaddress{Unter den Linden 6}
  \postcode{10099}
  \city{Berlin}
  \country{Germany} }
\affiliation{%
  \institution{Wirtschaftsuniversität Wien}
  \streetaddress{Welthandelsplatz 1}
  \postcode{1020}
  \city{Wien}
  \country{Austria} }
\affiliation{%
  \institution{Weizenbaum-Institut}
  \streetaddress{Hardenbergstr. 32}
  \postcode{10623}
  \city{Berlin}
  \country{Germany}}

\author{Henrik Leopold}
\email{Henrik.Leopold@klu.org}
\affiliation{%
  \institution{K{\"u}hne Logistics University}
  \streetaddress{Großer Grasbrook 17}
  \postcode{20457}
  \city{Hamburg}
  \country{Germany}}

 \author{Henning Meyerhenke}
\email{meyerhenke@kit.edu}
\orcid{0000-0002-7769-726X}
\affiliation{%
  \institution{Karlsruhe Institute of Technology (KIT)}
  \streetaddress{Am Fasanengarten 5}
  \postcode{76131}
  \city{Karlsruhe}
  \country{Germany}}

\author{Beno\^it Depaire}
\email{benoit.depaire@uhasselt.be}
\affiliation{%
  \institution{Hasselt University}
  \streetaddress{Martelarenlaan 42}
  \postcode{3500}
  \city{Hasselt}
  \country{Belgium}}

\renewcommand{\shortauthors}{Mendling et al.}
\newcommand{\suggest}[1]{\sethlcolor{green} \hl{#1}}
\newcommand{\del}[1]{\st{#1}}
\newcommand{\hmey}[1]{\suggest{#1}}

\begin{abstract}
\textbf{Abstract:} 
Research on algorithms has drastically increased in recent years. Various sub-disciplines of computer science investigate algorithms according to different objectives and standards. This plurality of the field has led to various methodological advances that have not yet been transferred to neighboring sub-disciplines. The central roadblock for a better knowledge exchange is the lack of a common methodological framework integrating the perspectives of these sub-disciplines.
It is the objective of this paper to develop such a research framework for algorithm engineering.  
Our framework builds on three areas discussed in the philosophy of science: ontology, epistemology and methodology. In essence, \emph{ontology} 
describes algorithm engineering as being concerned with algorithmic problems, algorithmic tasks, algorithm designs and algorithm implementations. \emph{Epistemology} describes the body of knowledge of algorithm engineering as a collection of prescriptive and descriptive knowledge, residing in World 3 of Popper's Three Worlds model. \emph{Methodology} refers to the steps how we can systematically enhance our knowledge of specific algorithms. The framework helps us to identify and discuss various \emph{validity concerns} relevant for any algorithm engineering contribution. 
In this way, our framework has important implications for researching algorithms in various areas of computer science. 
\end{abstract}



\maketitle
\citestyle{acmauthoryear}

\section{Introduction}
\label{sec:introduction}

The design, analysis, and evaluation of algorithms have been a major concern of computer science since its founding days. Algorithms constitute a specific class of technological rules~\cite{bunge1967search}. An \emph{algorithm} is a well-defined sequence of computational steps that transforms some input into some output~\cite{cormen2009introduction}. The number of steps, the effort they require, and the time they take have to be finite \cite{aho1974design}. Research on algorithms builds among others on foundational work by \citet{dijkstra1968constructive} and \citet{knuth1974computer}. The corresponding subdiscipline of computer science is called \emph{algorithmics}, of which \emph{algorithm engineering}~\citep{sanders2009algorithm} is a branch that embraces both theoretical and empirical contributions on algorithms and, thus, striving for a desirable symbiosis \citep{DBLP:journals/cacm/Ullman15,DBLP:journals/cacm/Mitzenmacher15}.

Various efforts have been made to build an inventory of algorithms developed through research in algorithms and algorithm engineering. Important works include the seminal book series by~\citet{DBLP:books/aw/Knuth68,DBLP:books/aw/Knuth69,DBLP:books/aw/Knuth73}, textbooks by~\citet{aho1974design,cormen2009introduction,harel2004algorithmics,sedgewick2013introduction}, and review books by~\citet{doi:10.1137/1.9781611976175,ferragina2023pearls}.
\citet{black2020dictionary} lists more than 300 algorithms, associated with 230 data structures, and 79 classic algorithmic tasks, covering areas such as automata, combinatorics, cryptography, memory management, geometry, graphs, numerical computation, parallel processing, quantum computation, searching, sorting, trees, and verification methods~\citep{black2020dictionary}. Though being extensive and covering seminal research, this list is far from complete. 
Today, researching algorithms is not only the focus of algorithmics and algorithm engineering, but also of various other sub-disciplines of computer science and neighboring research fields.
For 2024 only, Google Scholar lists more than 62,000 research papers with the term \emph{algorithm} in the title\footnote{\url{https://scholar.google.com/scholar?as_q=&as_epq=algorithm&as_oq=&as_eq=&as_occt=title&as_sauthors=&as_publication=&as_ylo=2024&as_yhi=2024&hl=de}.}, highlighting the importance and the richness of research on algorithms.

We observe that a review of the \textit{theory and practice of algorithm engineering} is missing. Much of methodological innovations are developed within the communities associated with 
specific conference series such as Very Large Databases (VLDB), Neural Information Processing Systems (NeurIPS), European Symposium on Algorithms (ESA), or Visualization (VIS), where they are partly fostered internally without being transferred to or adapted by neighboring communities. Is this pluralism an indication that there is no consensus in the field of computer science?  \citet{kuhn2012structure} calls such a consensus \emph{normal science} to signify the incremental work in an established paradigm, and interprets it as a sign of maturity of a field. However, the requirements, e.g., for evaluating algorithmic contributions differ quite considerably across the above-mentioned conferences: papers published at VIS put more emphasis on user studies, papers published at VLDB focus on performance evaluation and formal correctness. While there are good reasons for these differences, the question is whether a common scientific framework can integrate these differences in requirements, such that it might help other sub-disciplines of computer science to adopt methodological advances.

Against this background, we aim to explicate diverse perspectives on researching algorithms by the help of an integrated theoretical framework. To that end, we take foundational concepts discussed in the philosophy of science as a starting point. Our goal is to develop a theoretical framework along three dimensions, each related to a sub-discipline of the philosophy of science: ontology, epistemology, and methodology. In essence, \emph{ontology} is concerned with ``what is''. For algorithm engineering, this means we need to clarify what  the phenomena are that we consider. \emph{Epistemology} deals with the nature of knowledge about a specific phenomenon - i.e. ``what can we know about algorithms''. \emph{Methodology} refers to the study of method. It is concerned with the question ``how can we systematically enhance our knowledge'' of specific algorithms. With our framework, we aim to provide an answer to these questions and help researchers from different domains of computer science to learn from each other. In this way, our framework might eventually strengthen the \textit{cumulative tradition}~\citep{keen1980mis} in computer science, both within as well as across domains, such that researchers can easily build on each other's work and develop a repertoire of shared foundational concepts. 

This article is structured as follows.
Section~\ref{sec:ontology} presents the ontological perspective of our framework. Section~\ref{sec:epistemology} describes its epistemological perspective. Section~\ref{sec:methodology} discusses the methodological perspective. 
Section~\ref{sec:discussion} distills recommendations for algorithm engineering. 
Finally, Section~\ref{sec:conclusion} concludes our paper.

\section{Ontology of Algorithm Engineering}
\label{sec:ontology}
Ontology is the study that is concerned with the structure of the real world~\cite{bunge1977treatise,wand1990ontological}. An ontological theory of engineering clarifies the structure of real-world phenomena that relate to engineering.
In general, \emph{engineering} is defined as the practice of organizing the design, construction, and operation of any artifact that transforms the world around us for meeting a recognized need~\citep{rogers1983nature,vincenti1990engineers,staples2014-ontology}. \citet{staples2014-ontology} emphasizes that engineering deals with artifacts (algorithms in our context), that artifacts are meant to meet specific requirements, and that 
engineering builds on theories explaining why specific artifacts meet certain requirements.
The \emph{ontological perspective} of algorithm engineering refers to the entities associated with the practice of engineering algorithms. 

In turn, \emph{algorithm engineering} has been defined as a discipline that focuses on the design, analysis, implementation, tuning, debugging, and experimental evaluation of algorithms~\citep{demetrescu2004algorithm,sanders2009algorithm}. 
The salient feature of this definition is the term \emph{discipline}. It indicates that the design of algorithms is not the end of algorithm engineering, but the phenomenon that is studied. This means that its end is the generation of scientific knowledge about algorithms. While an algorithm design is an ontological entity in the realm of algorithm engineering practice, mind that knowledge about it is already an epistemological entity in the realm of the algorithm engineering discipline.
Also note that algorithms are often parts of larger~\emph{techniques} and \emph{systems}~\citep{munzner2014visualization}. These are more coarse granular entities, studied in neighboring disciplines such as systems engineering~\citep{sage1992systems}, information systems engineering~\citep{castro2002towards}, and design science~\citep{hevner2004design,peffers2007design,DBLP:books/sp/Wieringa14}.

Here, we build upon the ontological model of \citet{staples2014-ontology} and refine it for algorithm engineering. This model identifies ontological entities in a precise and explicit way according to principles generally accepted in computer science \cite{aho1974design} and operations research \cite{landry1983model}. These ontological entities have specific relationships with each other.

\begin{figure}[!htb]
    \centering
    \begin{minipage}{.6\textwidth}
        \centering       
\begin{enumerate}
    \item The first ontological entity is a \emph{real-world problem}, which resides in a specific problem context and which hints at an algorithm as part of an envisioned solution. 
    \item This real-world problem is in an abstraction relationship to an \emph{algorithmic task}. This task conceptualizes the real-world problem and captures its essential assumptions and requirements.
    \item Algorithmic tasks are in a satisfaction relationship with \emph{algorithm designs}. A design addresses the algorithmic task and incorporates design principles and design decisions. It can be characterized in terms of the performance guarantees it provides, e.\,g. derived from \emph{algorithm analysis}.
    \item An algorithm design is in an instantiation relationship with one or many \emph{algorithm implementations}. An implementation reflects implementation decisions and materializes the design. The execution of the algorithm implementation on data from the real-world problem generates \textit{results}. The implementation provides a specific output to the given real-world problem, as well as an indication of the algorithm's empirical performance. 
\end{enumerate}
    \end{minipage}%
    \hspace*{2mm}
    \begin{minipage}{0.37\textwidth}
 	\centering\includegraphics[width=0.62\linewidth]{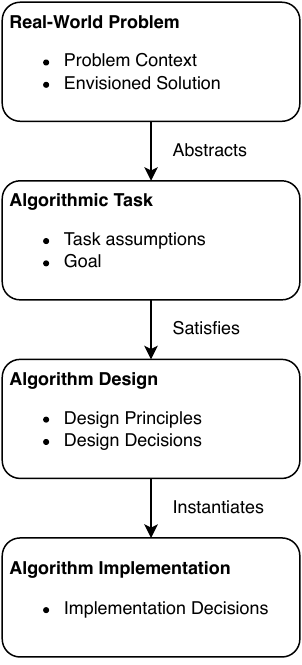}
     \caption{A framework for algorithm engineering: the ontological perspective}
 	\label{fig:ontology}
    \end{minipage}
    \vspace*{-1em}
\end{figure}

Figure~\ref{fig:ontology} provides an overview of our ontological framework. We discuss its four ontological entities in turn. For illustration proposes, we use examples from different areas of computer science including, among others, sorting, shortest-path identification, and natural language processing. 

\subsection{Real-World Problem}

The motivation for engineering algorithms can be found in real-world problems. Real-world problems are concrete, they are associated with dynamic situations and complex systems. They are often underspecified and their boundaries are fuzzy so that \citet{meyer2015nested} rather speak of a problem domain. \citet{ackoff1979future} emphasizes this aspect by referring to real-world problems as \emph{messes}. Real-world problems differ in terms of granularity, ranging from managerial problems to technical problems. 

\begin{example}
The textbook ``\textit{The Art of Computer Programming Volume 3: Sorting and Searching}" by \citet{DBLP:books/aw/Knuth73} describes several technical real-world problems that motivate the definition and analysis of sorting algorithms. ``\textit{If several files have been sorted into the same order, it is possible to find all of the matching entries in one sequential pass through them, without backing up.}''
\end{example}

\begin{example}
The textbook ``\textit{Tools for Thinking: Modelling in Management Science}" by~\citet[p.62]{pidd2003} describes several real-world problems, among others, a supermarket chain. This ``\textit{[...] supermarket chain must constantly have a full product offering on its shelves, must provide good customer service, must attract customers, must maintain low stocks and must make a profit. As people strive to achieve these aims, this throws up a continuing stream of issues through time. Examples might be the extent to which EPOS [Electronic Point of Sale] systems should be installed and used, or the question of how to get stocks from the manufacturer into the stores.}''
\end{example}

To solve a particular real-world problem, we can use an existing or a new algorithm design. This, however, requires to first translate the real-world problem into an algorithmic task.

\subsection{Algorithmic Task}
Many real-world problems expose themselves as deeply intertwined with the specifics of a particular problem domain. Modeling and analysis is required for structuring these problems, and various methods of management science and requirements engineering support such analysis~\citep{pidd2003,de2012requirements}. The result of problem-structuring is an \emph{abstraction} of the original real-world problem. Also the original messiness is resolved towards a precise formulation of what we call an \emph{algorithmic task}. Algorithmic tasks are also called \emph{algorithmic problems}, \emph{computational problems}, or just \emph{problems}~\citep{aho1974design}. Algorithmic tasks abstract from specific real-world problems, they represent \emph{classes of problems}~\cite{aken2004management}. The more abstract an algorithmic task is formulated, the broader is the extension of real-world problems it subsumes. Over time,
various algorithmic tasks have become standard tasks that are covered in textbooks, while new ones are continuously identified.

\begin{example}
\citet{DBLP:books/aw/Knuth73} describes several classical algorithmic tasks in his textbooks, among others sorting. He defines sorting as a specific task that considers as input a list of $n$ records and an ordering relation $<$ over the keys of these $n$ records. The goal of sorting is then to determine a permutation of the records that is consistent with the ordering relation~\cite[p.~5]{DBLP:books/aw/Knuth73}.
\end{example}

The example highlights that algorithmic tasks explicate \emph{assumptions} regarding the input and the processing of the input. The \emph{goal} for the algorithmic tasks specifies what the algorithm is meant to achieve in terms of output and performance. 
Also the supermarket chain might formulate its replenishment problem as an algorithmic task. Given the assumption that the input is a weighted graph without negative edges representing shops as vertices and roads as edges, the goal could be to find the shortest path in the graph covering all vertices in a minimum amount of processing time. Often, algorithmic tasks can be reformulated with different assumptions. Stronger assumptions can make tasks more specific. Also new algorithmic tasks are continuously introduced.

\begin{example}
\citet{black2020dictionary} defines the task of finding the shortest path as the ``\textit{problem of finding the shortest path in a graph from one vertex to another. "Shortest" may be least number of edges, least total weight, etc.}''
There is not just one algorithmic task for the shortest-path problem. \citet{deo1984shortest} describe a classification of shortest-path tasks, distinguishing usual path length and generalized path length with altogether 12 different variants. 
\end{example}

\begin{example}
\citet{bevilacqua2021recent} review contributions on word sense disambiguation. In essence, word-sense disambiguation algorithms map a word to its most likely sense in the context of a sentence or text~\cite{navigli2009word}. In their study, \citet{bevilacqua2021recent} point to related algorithmic tasks that have been defined in recent research. The word-in-context task requires as input two contexts to provide a prediction whether the same target words are used with the same meaning. The lexical substitution task disambiguates a word in context by finding semantically equivalent substitutes. The definition modeling task generates a description of the meaning of a word in context.   
\end{example}

In general, it is desirable that tasks are fully specified. In such a case, all required information can be provided as input and an algorithm determines the output fully automatically \citep{sedlmair2012design}. Sorting belongs to this category of tasks. Tasks can also be fuzzy in the sense that required input cannot be provided at the required level of precision.
In this case, interactive techniques can be designed that integrate several algorithmic components in such a way that a user can complement the output with background knowledge and human judgement \citep{meyer2015nested}. Such fuzzy tasks are often addressed by research in the field of computer visualization and natural language processing. For instance, word sense disambiguation is a fuzzy task since the semantics of natural language can hardly be fully specified. 
Ultimately, a key challenge is to specify the algorithmic task in such a way that an algorithm design is feasible.

\subsection{Algorithm Design}
Algorithm designs relate to algorithmic tasks. If appropriately developed, a design is in a \emph{satisfaction} relationship with a task. While the algorithmic task essentially describes \emph{what} is to be done and in which context (often by means of an input/output specification), the algorithm design specifies \emph{how} it works, that is, how to obtain the desired output. The act of designing considers the task's assumptions and yields a design specification that meets the goal of the task~\cite{DBLP:books/sp/Wieringa14}. The algorithm design implies that certain performance guarantees hold.

Designing is guided by knowledge~\citep{staples2014-ontology} on why algorithm designs meet specific goals. Design principles inform the design of novel algorithms. 
In essence, a {design principle} describes a solution strategy~\citep{gregor2020research}. Specific solution strategies such as divide-and-conquer, dynamic programming, backtracking, or local search~\citep{aho1974design,harel2004algorithmics,kleinberg2006algorithm} are available as a starting point for designing new algorithms. 

\begin{example}
A Vehicle Routing Problem (VRP) relates to the identification of the optimal route from one or several depots to several geographically scattered customers subject to constraints. \citet{laporte1992vehicle} discusses exact algorithms and heuristics to solve the VRP. He identifies three solution strategies among the exact methods - direct tree search, dynamic programming, and integer linear programming - and discusses various designs for each solution strategy. He also identifies common solution strategies for heuristic algorithms such as the nearest neighbor, insertion, and tour improvement procedure.
\end{example}

A \emph{design decision} is a decision with respect to the actual algorithm design that has the potential to change the behavior and performance of the design. Some design decisions are explicit, while others are implicit. In fact, every design can be expected to build on numerous implicit design decisions. For instance, when the algorithm is specified using Pascal-like pseudo code, this implies the assumption that the algorithm is likely to be implemented using a procedural programming language. This also entails assumptions about the machine executing the programming code~\cite{aggarwal1988input,sanders2009algorithm}. From a scientific perspective, mainly the explicit decisions are of interest as they have been made intentionally and can be expected to have the highest impact on the design. Such a design decision may involve using a particular design principle (i.e., a general proven solution strategy such as divide-and-conquer), paradigm (e.g., supervised learning or deep learning), or representation (e.g. word embeddings vs one-hot encoding). Note that the parameterization of specific design aspects is an explicit design decision, as it delegates the actual decision to the end-user. Instead of deciding at design time, a (hyper-)parameter is provided to control the actual behavior and performance.      

\begin{example}
     
In their paper ``\textit{Attention is all you need}'', \citet{vaswani2017attention} introduce the Transformer architecture. They explicitly discuss various design choices. Among others, they explain that building soley on the Attention mechanism is enough for the parallelization within training examples.
They also discuss that the use of Attention may lead to ``\textit{reduced effective resolution}'', which they counter by using Multi-Head Attention. Later in the paper, they devote an entire section (``\textit{Why Self-Attention}'') to demonstrate the impact of their design choices. 
\end{example}

Design decisions and the usage of design principles have implications for performance. Algorithm analysis can clarify these implications and associate algorithm designs with \emph{performance guarantees}
\citep{sanders2009algorithm,sedgewick2013introduction}. Such guarantees can be formulated as theorems over complexity classes.

\begin{example}
\citet{DBLP:books/aw/Knuth73} provides an overview of different sorting algorithms that use different solution strategies. The family of merge sort algorithms follows the design principle of divide-and-conquer. Its solution strategy is to decompose a problem into a hierarchy of partial problems that are partially solved in order to meet the goal of achieving an $\mathcal{O}(n \log n)$ run-time performance. 
\end{example}

\subsection{Algorithm Implementation}
\newtext{According to \citet{colburn2004methodology}, the ultimate goal of computer science is that programs can be efficiently implemented on physical systems.}
An algorithm design can be in an \emph{instantiation} relationship with several algorithm implementations. The implementation program closes the circle: it can be used as a concrete solution to address the original real-world problem by executing it with concrete input data on a specific computer. 

In practice, the implementation is actually hardly ever a simple instantiation~\cite{lukyanenko2020design}. 
Implementation is driven by implementation decisions. These refer to concrete \emph{implementation aspects} that address a specific goal based on an underlying rationale. 
For instance, the choice of a programming language dictates whether the operations of the algorithm's design can be directly translated. Furthermore, the runtime execution of the programming code inherits characteristics of the respective execution environment including hardware and system software characteristics. These depend upon memory hierarchy~\cite{DBLP:journals/eatcs/Sanders04}, caching strategies~\cite{karedla1994caching}, or how the compiler makes use of parallel processing capabilities provided by the processor~\cite{wolfe1996high}. 
Even if these factors are controlled, there are still a plethora of implementation options that can lead to performance differences of orders of magnitude.

\begin{example}
\label{ex:k-means-comparison}
Implementation decisions may have a drastic impact on algorithm performance, as evidenced by a study of~\citet{kriegel2017black}. The authors compare different implementations of the same algorithm designs. In their analysis of the reasons behind performance differences, they describe that certain implementations turned out be highly optimized. When re-implementing a k-means algorithm for a comparative evaluation, they ``\textit{chose the features with only eight dimensions [\dots] since k-means is known to work better in low dimensionality and we do not want to bias the analysis toward performance on high-dimensional data sets.}''
\end{example}

The execution of the algorithm implementation on data from the real-world problem generates two types of results. First, it produces an \textit{output}. For example, a web search algorithm generates a list of web sites that (desirably) contain relevant content related to the search query. Second, the algorithm implementation allows to obtain insights into the \textit{performance} of the algorithm. As for performance, we can identify two main dimensions: efficiency and effectiveness. Both can be captured by means of performance measures. Efficiency refers to the processing performance, typically measured in terms of time and memory consumption. Effectiveness, on the other hand, concerns the quality of the output in relation to the real-world problem. Depending on the problem at hand, a wide range of measures exist. In the context of web search, we could, for instance, employ metrics from information retrieval such as precision, recall, and F1-measure~\citep{baeza1999modern}. To what extent the achieved performance of the algorithm is considered sufficient depends on the goals that were defined for the algorithmic task and is, hence, highly context-dependent. In web search, one could imagine that the goal is to outperform an established baseline, such as Google search, with respect to the chosen performance metrics.   

\begin{example}
    \citet{devlin2014fast} present a new neural network joint model (NNJM) for statistical machine translation. Their core idea is to determine a source context window, i.e, the number of words from the source language that are relevant to find the correct word of the target language. To provide insights into the \textit{performance} of the respective implementation, the authors conduct extensive experiments on Arabic-English and Chinese-English data sets. They show that their approach is superior in comparison to existing baselines both in terms of \textit{effectiveness} and \textit{efficiency}. For quantifying efficiency, they use the metrics \textit{lookups/second} and \textit{seconds/word}. For quantifying \textit{efficiency}, they use the \textit{BLEU} score, an established metric for measuring the similarity of machine-translated text given a reference translation.    
\end{example}

\section{Epistemology of Algorithm Engineering}
\label{sec:epistemology}
The previous elaborations emphasize that the practice of algorithm engineering 
is concerned with finding an algorithm implementation for a given real-world problem. The science of engineering takes a more abstract view. It is not about solving individual problems, but about ``\textit{general knowledge, linking an intervention or artifact with a desired output or performance
in a certain field of application}'' \cite{aken2004management}. 
Thus, algorithm engineering aims to expand the body of knowledge on algorithm design.
The \emph{epistemological} perspective of our framework addresses questions of what we can know about an algorithm. 
We use Popper's three worlds as a conceptual model to clarify the relationship between knowledge and algorithms. Note that we are not the first ones to build on Popper's model. It has been used for clarifying general concepts of programming by~\citet{naur1985programming} and of engineering by~\citet{staples2014-ontology}. We describe that the body of knowledge contains knowledge \textit{of} and \textit{about} tasks as well as knowledge \emph{of} and \textit{about} designs. 

\begin{figure}
 \centering\includegraphics[width=1\linewidth]{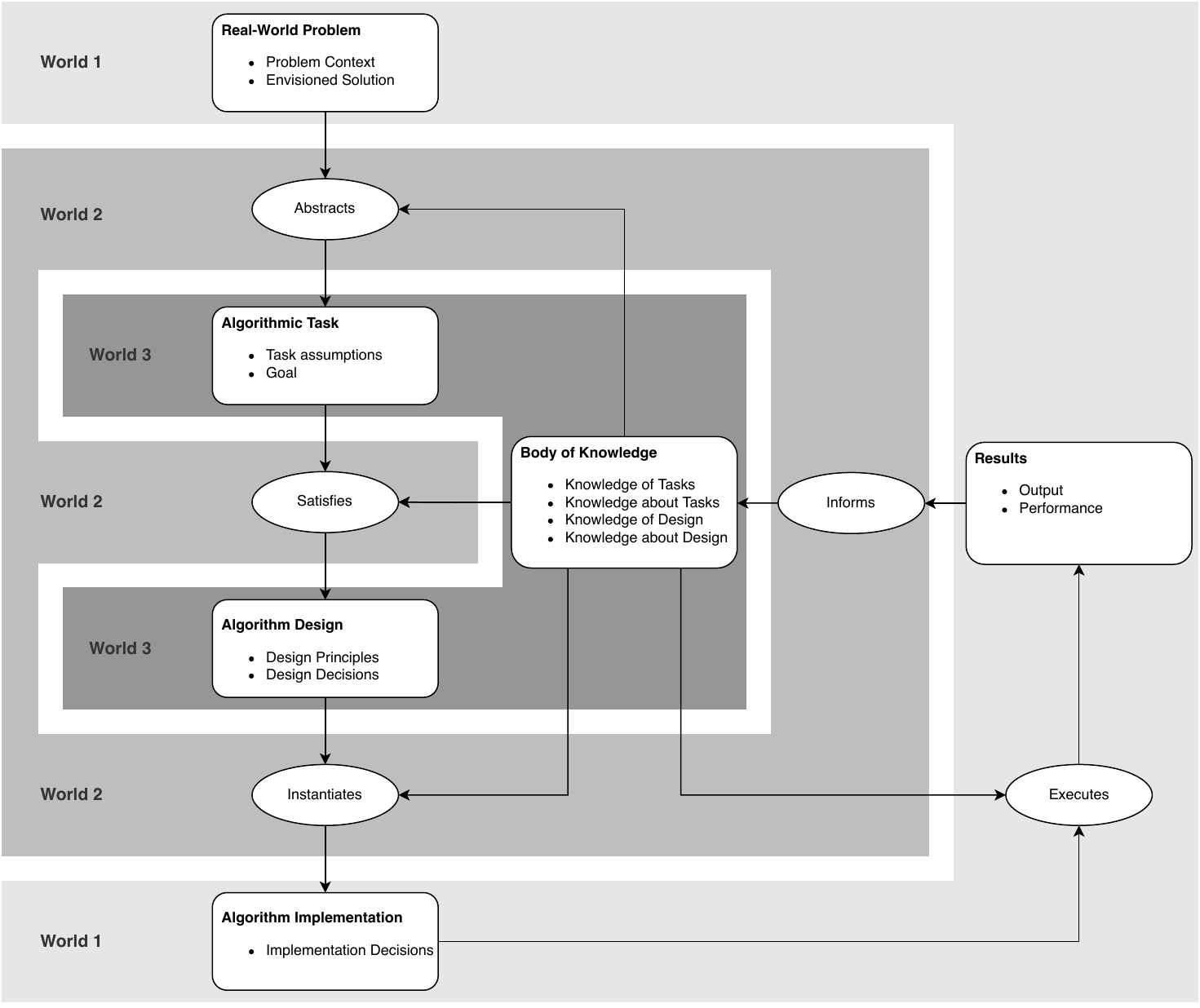}
 \caption{A framework for algorithm engineering: the epistemological perspective}
 \label{fig:epistemology}
\end{figure}

\subsection{Popper's Three Worlds}

The body of knowledge of algorithm engineering relates to different spheres that \citet{popper1979three} describes in his epistemological model of three worlds. In essence, he distinguishes World~1 of physical entities, World~2 of subjective mental states, and World 3 of objective knowledge. World~1, often referred to as the real world, is assumed to exist independently from an observer and to behave in a regular way. The goal of science is to formulate objective, empirical knowledge in World~3 that refers to World~1. Such scientific knowledge is created by humans, whose individual mental states define World~2, mediating between the other two worlds.

Figure~\ref{fig:epistemology} extends our ontological perspective on algorithm engineering and shows how it relates to Popper's three worlds. Our discussion builds on arguments by \citet{naur1985programming} and \citet{staples2014-ontology}. First, both real-world problems and algorithm implementations are entities that exist in the physical world independently from an observer; they belong to World~1. 
Second, the three actions abstract, design, and implement build on human cognition and mental models. For this reason, they reside in World~2. Third, the algorithmic tasks, algorithm design, and the body of knowledge are part of World~3 as these are meant to be objectively described. Regarding the body of knowledge, we distinguish between knowledge related to tasks and knowledge related to designs. For both types we further differentiate between \textit{knowledge of} and \textit{knowledge about}. In the context of tasks, \textit{knowledge of} relates to \textit{what is} a specific task, while \textit{knowledge about} relates to \textit{what are properties} we know about these tasks. In a similar way, knowledge of design relates to \textit{what is} a specific algorithm design, while \textit{knowledge about} relates to \textit{what are properties} we know about this algorithm design.  

Popper's three world view unveils an important implication for algorithm engineering that is easily overlooked: Any scientific knowledge resides in World~3. The same observation applies to algorithm engineering and its body of knowledge. Mind that our understanding of a specific real-world problem or a specific algorithm implementation is ontologically different: both reside in World~1. World~1 is what matters from a practical perspective, whereas World~3 is what we are interested in from a scientific perspective. In other words, from a practical perspective we are interested in real-world problems and algorithm implementations, whereas from a scientific perspective we focus on knowledge of and about algorithmic tasks and designs.
We now discuss in turn what knowledge of and about tasks as well as knowledge of and about algorithmic designs can be formulated.

\subsection{Knowledge of Tasks}

Real-world problems reside in World 1 and are often ``\textit{wicked problems}''~\cite{rittel1973dilemmas}. 
\newtext{This wickedness implies that such problems are unique, having no definitive formulation, no explicit stopping rule, no ultimate test of solution, and affording numerous alternative explanations.}
Their wicked nature is the reason why algorithmic tasks are not connected with a problem via a simple isomorphic mapping. \newtext{According to \citet{rittel1973dilemmas}, they have to be ``tamed'' by constructing an abstraction.} We refer to the outcome of this abstraction process as \textit{knowledge of tasks}. Knowledge of tasks, therefore, captures what tasks exist and how they are structured. As pointed out earlier, keep in mind that our notion of an algorithmic task is also referred to as \emph{algorithmic problem}, \emph{computational problem}, or just \emph{problem} in the literature~\citep{aho1974design}. These are tasks (residing in World 3) and should, however, not be confused with our notion of a real-world problem (residing in World 1). 
Our notion of algorithmic tasks and World 3 problems are semantically equivalent. 
To avoid confusion, we will use the term \textit{(algorithmic) task} throughout this paper. For established tasks, such as the \emph{traveling salesperson problem}, we will, however, adopt the original term. 

Depending on the context, tasks and their structure might be described mathematically, formally, semi-formally, or also informally.    

\begin{example}
\label{ex:tsp}
The \emph{traveling salesperson problem} can be described mathematically, modelled as an optimization problem.
\citet[p.104]{hromkovivc2013algorithmics} defines it as follows:
\begin{itemize}
    \item \textit{Input}: a weighted complete graph $(G,c)$ where $G=(V,E)$ and $c: E \mapsto \mathbb{N}$. Let $V=\{v_1, \dots, v_n \}$ for some $n\in \mathbb{N}-\{0\}$.
    \item \textit{Constraints}: For every input instance $(G,c)$, $\mathcal{M}(G,c) = \{ \langle v_{i_1}, v_{i_2}, \dots, v_{i_n}, v_{i_1}\rangle~|~(i_1,i_2,\dots,i_n$ is a permutation of $(1,2,\dots,n)\}$, i.e., the set of all Hamiltonian cycles of $G$.
    \item \textit{Costs}: For every Hamiltonian cycle $H= \langle v_{i_1}, v_{i_2}, \dots, v_{i_n}, v_{i_1}\rangle \in \mathcal{M}(G,c)$, the cost is induced by the cost function $cost(\langle v_{i_1}, v_{i_2}, \dots, v_{i_n}, v_{i_1}\rangle, (G,c)) = \sum^n_{j=1} c(\{ v_{i_j},v_{i_{(j\mod{n})+1})}\})$. 
    \item \textit{Goal}: Minimum.   
\end{itemize}
\end{example}

\begin{example}
\citet{van2011process} formally introduces the task of \emph{process discovery} in the context of process mining. Informally, this task can be described as identifying a function that constructs a process model from an event log (i.e., a set of execution sequences) such that the model is representative for the behavior observed in the event log. Formally, he first clarifies the notion of an event log: A simple event log $L$ is a multi-set of traces over a set of activities $\mathcal{A}$, i.e., $L \in \mathcal{B}(\mathcal{A}^{*})$. Building on this definition, he then formally describes the process discovery task as identifying a function $\gamma$ that maps an event log $L \in  \mathcal{B}(\mathcal{A}^{*})$ onto a Petri net $N$ with a marking $M$, i.e., $\gamma (L) = (N,M)$. It is important to note that such formal task descriptions often contain text. As a result, the degree of formality can vary and the transition to fully informal task descriptions (see example below) is often fluid.   
\end{example}

\begin{example}
\citet{russakovsky2015imagenet} list several \emph{image recognition} tasks in relation to ImageNet. They use an informal, textual description to specify the \emph{object detection} task. The goal of an algorithm addressing this task is to produce bounding boxes at the right position and scale for all instances of a given object category. The effectiveness of algorithms is evaluated by recall (the share of correctly detected target object instances) and precision (the share of spurious detections).
\end{example}

\subsection{Knowledge about Tasks}
\emph{Knowledge about tasks} relates to insights about properties of specific algorithmic tasks. The analysis of tasks can yield formal knowledge and empirical knowledge.

\emph{Formal knowledge} has been established for a wide range of tasks. These tasks are often referred to as computational problems. A key concern in the analysis of tasks is the question whether a problem is tractable. An \emph{intractable problem} can be solved in theory, but any algorithm in practice would take too much time or space on inputs of non-trivial size~\citep{hopcroft2000introduction}. 

Complexity classes have been identified for specific tasks. A hierarchy of these classes ranges from non-deterministic with logarithmic memory space \textbf{NL} to exponential space \textbf{EXPSPACE}. Many classical problems have been identified as members of the class that can be solved in polynomial time on a non-deterministic Turing machine~\textbf{(NP)}~\citep{garey1979computers}.
An important subclass is the set of \textbf{NP-complete} problems;
loosely speaking, its members
are at least as difficult to solve as any other problem in \textbf{NP}.
A more fine-grained analysis allows to reason about fixed-parameter
tractable problems \textbf{(FPT)}, which can be solved in time $f(k) \cdot |(x,k)|^{O(1)}$ for some
computable function $f$, where $k$ is some parameter of the input~\citep{cygan2015parameterized}.

\emph{Empirical knowledge} is of practical interest in particular for problems that have been identified as intractable.
Key to this knowledge is the identification of patterns in typical input data that can be exploited for the design of algorithms. Improvements on many algorithmic tasks have been inspired by the availability of \emph{benchmark data sets}. These data sets are publicly shared data sets of realistic input of a specific algorithmic task.

\begin{example}
A classical algorithmic task that is known to belong to the class of \textbf{NP-complete} problems is the traveling salesperson problem (see \autoref{ex:tsp}). It assumes a list of cities and distances between them. In its optimization
version, the question is \emph{what is the shortest route that visits all cities and returns to the origin?} 
(Its decision version asks if such a route of a specified length or less exists.)
Formal knowledge on this task has been established among others by~\citet{papadimitriou1977euclidean}, who proves that the Euclidean travelling salesperson problem is \textbf{NP-complete}.
Major breakthroughs on TSP have been inspired by the availability of realistic problem instances and corresponding benchmark data sets~\citep{sanders2009algorithm}; such as the 8th DIMACS Implementation Challenge\footnote{http://dimacs.rutgers.edu/archive/Challenges/TSP/} or the World TSP Tour.\footnote{\url{https://www.math.uwaterloo.ca/tsp/world/}.} By help of these data sets, empirical knowledge about properties of the input data could be established and exploited, for instance by considering power-law distributions~\citep{ouaarab2014discrete}. In many domains of algorithm engineering, data sets such as those for TSP~\citep{drori2020galaxytsp} are continuously extended to challenge improvements and innovation of algorithms and knowledge about tasks.
\end{example}

\begin{example}
The understanding of algorithmic tasks in the field of image recognition is driven by the availability of large-scale annotated image data sets such as ImageNet~\citep{russakovsky2015imagenet}. ImageNet contains several million images that have been manually annotated with names of objects. Research on ImageNet has not only led to major advancements of algorithms for image recognition, but also of techniques for systematically annotating a large amount of images using crowdsourcing~\citep{russakovsky2015imagenet}.   
\end{example}

\subsection{Knowledge of Design}
\emph{Knowledge of design} refers to what an algorithm design is, how it works, and how it addresses an algorithmic task.
Such knowledge of design is \emph{prescriptive}, as enables the implementation of concrete algorithms by humans. \citet{gregor2013positioning} refer to it as $\Lambda$ knowledge. The body of knowledge that relates to design largely consists of a repertoire of \emph{established algorithm designs} that have demonstrated their efficiency and effectiveness in prior studies, i.e. where formal or empirical knowledge indicates that an algorithm design meets the goals defined for the algorithmic task.   
Established algorithm designs range from concrete designs that address specific tasks to general procedures that are applicable for a wide range of tasks. Regardless of the scope, algorithm designs may build on generic \textit{design principles}, which describe general ideas for finding or constructing a solution for a given algorithmic task. Among others, \citet{harel2004algorithmics} as well as \citet{kleinberg2006algorithm} describe several examples of such principles including divide-and-conquer. To illustrate the broad range of algorithm designs, consider the following four examples.  

\begin{example}
   \citet{grund2021codeshovel} introduce an approach that identifies earlier versions of a considered source code method in the file history. The central artifact of this approach is a specifically designed algorithm that addresses this 
   specific task in the context of software engineering and source code management.  
\end{example}

\begin{example}
\citet{gevay2021efficient} introduce a novel distributed dataflow system called ``\textit{Mitos}'' that allows to efficiently coordinate the distributed execution of control flow in a user-friendly fashion. The proposed system consists of several different components that are also separately introduced. Some parts are presented at the algorithmic level, while others are discussed conceptually. As for the scope, the system can be used for a wide range of analysis tasks that benefit from distributed execution.
\end{example}

\begin{example}
   \citet{bentley1980multidimensional} introduces an approach that targets problems dealing with collections of objects in a multidimensional space. The author positions his approach as a generic algorithmic design 
   as it can be instantiated and used to  give best-known solutions to several problems including range searching, closest pair, and nearest neighbor.
   For a more general setting, \citet{bentley1999programming} presents a set of established and frequently used algorithm design techniques for solving algorithmic tasks in practice.
\end{example}

\begin{example}
In a recent collection on important design techniques,
\citet{ferragina2023pearls} selects algorithmic tasks that admit
``\textit{surprisingly elegant [design] solutions that can be described in a few lines of code}''.
The topics covered are numerous and include random sampling, list ranking, 
sorting, string search, and compression. The variety of topics is also reflected
by the variety of algorithm design principles presented.
\end{example}

Knowledge of what a \emph{concrete algorithm design} is can be explicated in different ways. At least three different levels can be distinguished based on their proximity to the implementation. First, the algorithm's design can be described using pseudocode. Pseudocode is more abstract and more compact than code and yet typically facilitates the re-implementation of the design in various programming languages. Tool support is available for generating pseudocode for actual implementations \citep{oda2015learning}. Second, diagrams such as flow charts can be used to describe the algorithm design. Such diagrams provide benefits in terms of cognitive effectiveness~\citep{malinova2021cognitive} and they have been found to be superior in terms of comprehension in comparison to pseudocode~\citep{scanlan1989structured}. 
Third, algorithms can be described using formal specifications, e.g. by mathematical formulae.  
It is worth noting that not every type of specification is equally useful for any type of design. While specific algorithms can be well described using pseudocode or formal specifications, this does not apply to larger systems consisting of several algorithmic subcomponents. Below, we discuss a number of examples to illustrate how different specifications are used.    

\begin{example}
\label{ex:spec_pseudocode}
It is customary in many algorithm engineering papers and in most textbooks
on the topic (see e.g., \citet{ferragina2023pearls,DBLP:books/sp/SandersMDD19}) to use pseudocode.
\citet{grund2021codeshovel} also use pseudocode to describe how the proposed approach identifies earlier versions of a considered source code method in the file history.  
\end{example}
\begin{example}
\label{ex:spec_flowchart}
  \citet{Chen2020AML} use a flowchart for clarifying the data processing pipeline of their system. This system builds on a machine-learning-based approach that helps detecting vulnerabilities in  open-source libraries.  
\end{example}
\begin{example}
\label{ex:spec_formal}
    \citet{zhou2021informer} use mathematical equations as formal specifications to describe their method for long sequence time series forecasting. They also use this specification as a basis to prove relevant properties of the proposed method.  
\end{example}

\subsection{Knowledge about Design}

\emph{Knowledge about design} relates to insights about properties and characteristics of the respective algorithm design. The relationship between algorithmic tasks and designs is often explicated as specific \textit{hypotheses}. In their deductive-nomological model, \citet{hempel1948studies} describe a hypothesis as a law bound to a set of conditions. This follows the structure of an \textit{explanans} explaining an \textit{explanandum}, a specific observation. For algorithm engineering, the explanandum is often the extent to which an algorithm \emph{satisfies} the requirements defined by the algorithmic task, the explanans refers to the design principles realized by the algorithm design and the assumptions made by the algorithmic task \citep{simon1969sciences,staples2014-ontology,DBLP:books/sp/Wieringa14,hall2017design}. The algorithmic task and the algorithm design directly correspond to our ontological framework described earlier in Figure~\ref{fig:ontology}. Knowledge about design can be substantiated in a formal and in an empirical way.

\begin{example}
To gain formal knowledge on the performance of an 
algorithm design, \citet{ferragina2023pearls} uses (among others) the external-memory (EM) model for analysis. 
This model focuses on the I/O bottleneck in computations by assuming a two-level memory hierarchy
with a fast (yet small) internal memory and a large (yet slow) external memory. 
Data is retrieved from the external memory in chunks, which mimics block transfers in real systems.
In this way, as a ``\textit{golden rule}''~\cite[Ch.~1]{ferragina2023pearls}, algorithm designs are steered towards exploiting locality.
\end{example}

\begin{example}
\label{ex:hypothesis_formal}
  \citet{karlin2021slightly} introduce a new approximation algorithm for the travelling salesperson problem. They use theorems and proofs to show that, given a constant $\epsilon$, their algorithm delivers a solution that has at most $\frac{3}{2}-\epsilon$ times the cost of the optimal solution.     
\end{example}
\begin{example}
\label{ex:hypothesis_empirical}
\citet{gevay2021efficient} use a plot to demonstrate how the execution time of their proposed system ``\textit{Mitos}'' changes with increasing size of the input.
\end{example}

There are different \textit{types} of knowledge about design. Following work by~\citet{santner2003design} on the design and analysis of computer experiments, we distinguish four general knowledge types that are often investigated in the field of algorithm engineering: performance, sensitivity, uncertainty, and explanatory knowledge. 

\emph{Performance knowledge} focuses on the algorithm's performance in terms of the extent that the algorithm \emph{meets its task requirements}.  Performance knowledge can be formulated as a question of satisfaction (Does the algorithm design \textit{satisfy} the task requirements?) or as a question of degree (To which extent does the algorithm design satisfy the task requirements \textit{better} than other designs?) \citep{staples2014-ontology,DBLP:books/sp/Wieringa14,hall2017design}.
\newtext{Often, knowledge claims on performance refer to general criteria~\citep{larson2025} or specific measurements.}
Research focused on performance knowledge often relies on a competition-based evaluation or formal proofs against state-of-the-art algorithms. 

\begin{example}
\label{ex:performance_faster}
    \citet{gevay2021efficient} introduce the system ``\textit{Mitos}'' for task analysis. The key feature of the proposed system is that it is considerably faster than the state of the art.  
\end{example}
\begin{example}
\label{ex:performance_better}
    \citet{kuccuk2021improving} introduce an approach for statistical fault detection in source code. The key feature of the proposed approach is that it performs considerably better than the state of the art with respect to fault localization. 
\end{example}

\emph{Sensitivity knowledge} focuses on how robust the performance of an algorithm is in face of changes of \emph{internal} design decisions. These design decisions are often parameterized and the goal is to evaluate the robustness of the algorithm's performance against sub-optimal parameter settings. 

\begin{example}
\label{ex:sensitivity}
\citet{Chen2020AML} propose a machine-learning-based approach that helps detecting vulnerabilities in open-source libraries. To this end, they build on word embeddings from word2vec. Recognizing that word2vec relies on a number of parameters (e.g., window size and vector size), the authors systematically explore how the performance of the approach is affected by changing word2vec-related parameters for different data sets. They find that the parameters that predominantly affect the performance differ from data set to data set.   
\end{example}

\emph{Uncertainty knowledge} considers the performance of an algorithm as a function of the task assumptions, which describe the relevant aspects of the \emph{external} environment of an algorithm. 
\newtext{\citet{larson2025} use the term context claims to refer to such knowledge.} The goal of uncertainty analysis is to assess how the expected performance of an algorithm varies across problem instances. Alternatively, the goal could also be to investigate the precision of the expected performance.

\begin{example}
     \citet{kuccuk2021improving} propose an approach for fault localization in source code. In the evaluation experiments, the authors do not only show that the approach is performing better than the state of the art, they also investigate the impact of a specific data set characteristic called \textit{covariate imbalance}. They analyze the empirical relationship between covariate imbalance and the chosen performance metric and find that increasing covariate imbalance indeed leads to higher fault localization costs. They also show, however, that the presented technique is less affected by a covariate imbalance than the state of the art.
\end{example}

\emph{Explanatory knowledge} provides insights into the \emph{mechanisms} of how task assumptions and design decisions interact and influence the algorithm's performance. 
Key for establishing such knowledge is the comparative evaluation of different configurations of design, for instance, by including and excluding certain functionality, and then studying the effect on results.

\begin{example}
\label{explaexrah related to Santner as :explanatory}
\citet{he2016deep} present a residual learning framework for image recognition to ease the training of networks that are substantially deeper than the state of the art. In their experiments, the authors analyze edge cases, such as a network with more than 1000 layers. They find that the performance of such extremely deep networks is not as good as the performance of networks with around 100 layers. They conclude that this is caused by overfitting and that networks with more than 1000 layers are unnecessarily large.   
\end{example}

Research papers differ in the extent to which they emphasize \emph{knowledge of} and \emph{knowledge about} design. 
There are papers that focus on knowledge of design. A paper proposing a new algorithm is a typical example of this kind. Mind, however, that such papers often include a larger evaluation part that essentially provides knowledge about design, in which  properties of the algorithm are analyzed to justify its underlying hypotheses. There are also papers offering knowledge about design only. Survey papers, papers on new theoretical insights about existing algorithms, or benchmarking studies belong to this category. These studies often do not propose new algorithms, but compare or analyze previously published ones.

\begin{example}
\label{ex:benchmark}
\citet{wang2021we} address the problem of cardinality estimation  in the context of query optimization. More specifically, they set out to answer the question whether learned models for cardinality estimation are ready for use in practice. To answer this question, the authors conduct a comprehensive study with five learned methods and eight traditional methods for cardinality estimation and four real-world data sets. The study consists of three different parts and provides deep insights into the performance of the studied methods in different environments and when learned methods fail to deliver correct results. 
\end{example}

\begin{example}
    \citet{10.1145/2027216.2027217} provide theoretical evidence on why the
    very popular $k$-means clustering algorithm from the 1950s is usually very fast in practice -- despite its provably exponential worst-case
    running time. They show, based on smoothed analysis, that the smoothed number
    of iterations is in fact bounded by a polynomial in the input size and the 
    (inverse of) the standard deviation of a Gaussian perturbation of the data.
\end{example}

\newtext{In the end, it is important to recall that algorithmic designs as mathematical abstractions and algorithm implementations as physical programs are epistemologically different~\citep{colburn2004methodology}. Contributions to the formal verification debate in the 1970s and 1980s argued 
that program verification was neither practically~\citep{de1979social} nor philosophically feasible~\citep{fetzer1988program}. Abstractions of computer languages, procedures and data have become so firmly established in computer science that many agree with \citet{hoare1985mathematics} who posits the equivalence of computer programs and mathematical expressions as much as programming languages and mathematical theory. However, the correspondence between physical systems executing algorithms and mathematical specifications is not self evidence and has to be firmly established: ``An axiom system may just happen to describe physical reality, but that is for experimentation in science to decide''~\citep{colburn2004methodology}. For that reason, both formal and empirical knowledge are equally important for advancing algorithm engineering.}

\section{Methodologies of Algorithm Engineering}
\label{sec:methodology}
We emphasized above that the general goal of algorithm engineering is to extend the body of knowledge of algorithm designs, algorithmic tasks, and corresponding knowledge. Up to this point, however, we have focused on \textit{what} we can know. In this section, we elaborate on the question of \textit{how} to systematically extend the body of knowledge. 
We identify four categories of such extensions:
\begin{enumerate}
    \item 
    New or better knowledge \emph{of tasks}.
    \item 
    New or better knowledge \emph{of designs};
    \item 
    New or better formal knowledge \emph{about tasks and designs};
    \item 
    New or better empirical knowledge \emph{about tasks and designs}.
\end{enumerate}
Many papers present extensions of the mode knowledge that are 
\emph{improvements} of one or the other kind~\citep{gregor2013positioning}. For instance, an improvement could be a new sorting algorithm that performs better than existing ones, a new theorem that identifies narrower bounds than earlier published ones, or more specific empirical results obtained by an experiment.
Some papers develop knowledge on how existing solutions can be transferred to problems that have not yet been discussed. \citet{gregor2013positioning} call this \emph{exaptation}.
Such knowledge extensions might be an algorithm that solves a newly defined algorithmic task, a theorem on an algorithm for which no theorems had been proven, or a first empirical study on an algorithm.

\begin{figure}
 \centering\includegraphics[width=\linewidth]{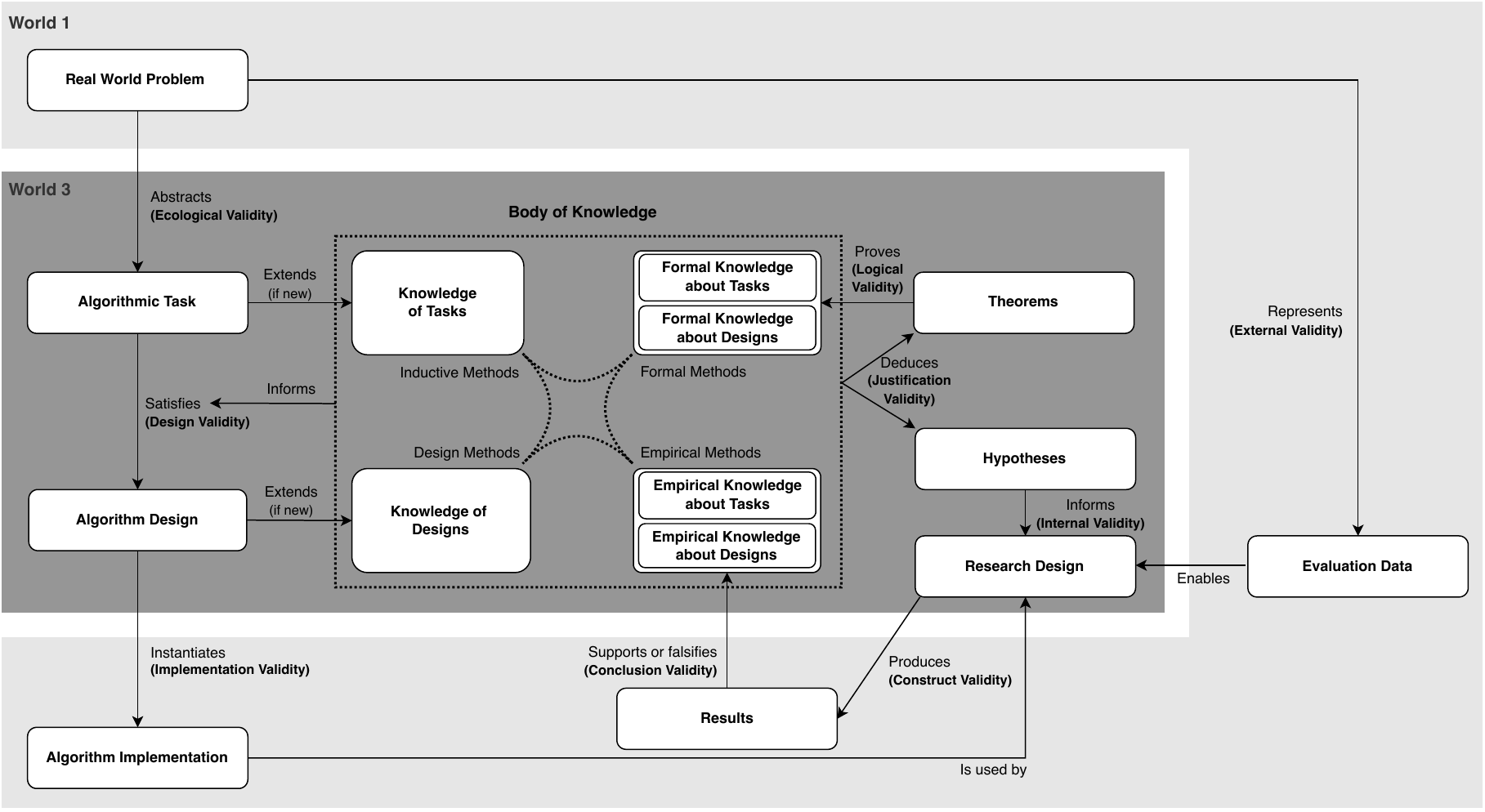}
 \caption{A framework for algorithm engineering: the methodological perspective}
 \label{fig:methodology}
\end{figure}

Figure \ref{fig:methodology} summarizes the methodological perspective of our framework. There are two key observations. First, additions to the body of knowledge  
are at the center of iterative processes~\cite{sanders2009algorithm,DBLP:books/sp/Wieringa14}. 
New algorithmic tasks and algorithm designs extend the corresponding knowledge categories; new theorems and corresponding proofs establish formal knowledge about algorithms; and new empirical results inform empirical knowledge about algorithms.
Second, each knowledge type is associated with corresponding research methods: inductive methods, design methods, formal methods\footnote{
Here and in the following, when we refer to \emph{formal methods}, we mean
a broad set of rigorous theoretical techniques that allow to derive formal 
knowledge about algorithmic tasks and/or algorithm designs. 
Due to a different focus, this is to be distinguished from the widespread 
usage of the term for mathematically rigorous techniques and tools employed for the specification, 
development, analysis, and/or verification of software and hardware systems.
}, and empirical methods. 
\newtext{ Validity concerns are of specific importance for methods.
Table \ref{tab:validity} summarizes the key validity concerns for algorithm engineering. In the following, we describe each method in turn and highlight how these concerns inform and shape their application.} 
\begin{table}[bth]
\footnotesize
    \centering
    \begin{tabular}{lp{6cm}l}
      \toprule\noalign{\smallskip}
         \textbf{Validity concern} & \textbf{Explanation} & \textbf{Affected
         entities} \\\noalign{\smallskip}
         \toprule
         Ecological validity & The extent to which an algorithmic task or setup reflects real-world conditions and problem contexts (based on~\citet{ashcraft2010cognition,holleman_real-world_2020}). & Algorithmic task\\\noalign{\smallskip}
         
         Design validity & The degree to which the internal structure and logic of an algorithm design is coherent, justified, and explainable~\cite{larsen2020validity}. & Algorithm design\\\noalign{\smallskip}
         
         Implementation validity & The extent to which an algorithm implementation faithfully instantiates the intended design and behaves as expected (based on~\citet{lukyanenko2015guidelines,lukyanenko2020design}). & Algorithm implementation\\\noalign{\smallskip}

         External validity & \newtext{The degree to which results generalize across data sets of interest (based on~\citet{cook2002experimental}).} & Empirical results \\

         Justification validity & 
         \newtext{The degree how convincingly a hypothesis or theorem is supported by a deductive argument (based on~\citep{lannin2005generalization}).}
         & Theorems, hypotheses  \\\noalign{\smallskip}

         Logical validity & 
         \newtext{The degree to which the syllogisms used in a proof preserve truth (based on Aristotle and reflection in~\citet{durand2008truth}).}
         & Proofs  \\\noalign{\smallskip}

         Internal validity & The extent to which observed effects can be attributed to the treatment rather than to confounding factors~\citep{wohlin2012experimentation}. & Research design \\\noalign{\smallskip}
         
         Construct validity & The degree to which the measure of a construct accurately measures the intended property~\citep{o1998empirical}. & Measurement \\\noalign{\smallskip}
         
         Conclusion validity & \newtext{The degree to 
          which the results can reasonably be regarded as revealing the hypothesized connection~\cite{cook1979design,garcia2012statistical}.}
         & Empirical results\\\noalign{\smallskip}
         \bottomrule
    \end{tabular}
    \caption{Validity concerns in algorithm engineering}
    \label{tab:validity}
\end{table}

\subsection{Methods for Generating Knowledge of Algorithmic Tasks}\label{know-of-task-methods}
We already highlighted that the wicked nature of real-world problems is the reason why algorithmic tasks are not a simple isomorphic mapping of the problem and, therefore, need to be constructed. Various methods have been developed and used in operations research~\citep{pidd2003,mingers2004problem}, requirements engineering~\citep{bourque2014guide}, and visual analytics~\citep{munzner2014visualization} for creating a specification of a real-world problem. Essentially, any research method that operates inductively on qualitative empirical data such as case study methods, focus group designs, interview studies, or think-aloud techniques~\citep{recker2012scientific} can be applied.  

\begin{example}
\citet{wu2022rasipam} develop techniques for interactive pattern mining and corresponding algorithms. They followed the design study methodology by~\citet{sedlmair2012design} for analyzing the real-world problem domain of racket sports. They interview domain experts, formalize their analysis tasks, and involve them in the design and evaluation of a prototype. In this way, they identify five analysis tasks and corresponding constraints. The resulting task descriptions inform the design of the author's interactive tactics mining algorithm. Input to this algorithm are raw sequence data of racket matches. 
\end{example}

\emph{Models} are often used for specification in general and for algorithmic tasks more specifically. In this context, a model is defined as (i) a mapping from some original (ii) by means of an abstraction operation (iii) in order to serve a specific purpose~\citep{kuhne2006matters,mendling2008metrics}. Various methods can be used for constructing models that structure problems. For an overview see ~\citet{mingers2004problem}. 
\begin{example}
Algorithms play an important role in supporting supply chain operations.
\citet{biswas2004object} provide an extensive analysis of supply chain tasks with the ambition to devise algorithmic decision support. They use different UML models: class diagrams for clarifying a) the relationship between strategies and policies and b) the objects of a supply chain, as well as activity diagrams for specifying a) the steps of addressing the location problem and b) the behaviour of the dealer object.
Beyond that they use informal models to represent concept taxonomies and their proposed system architecture. 
\end{example}

\emph{Examples} of real-world problem instances also play an important role for describing tasks. 

\begin{example}
\citet{agarwal2018efficient} analyze the problem of path planning amid a set of obstacles in a two-dimensional plane. Their work is motivated by robot motion planning used e.g. for surgeries. They formalize this real-world problem to a task of constructing an approximate minimal-cost path algorithm. They illustrate the task using several examples of Voronoi diagrams.
\end{example}

The concern of \emph{ecological validity} is of particular importance for the specification of tasks. It refers to what extent the task characteristics relate to the real world such that results will generalize to real-world problems \cite{ashcraft2010cognition,holleman_real-world_2020}. This is in essence a concern that touches upon the relationship between a class of real-world problems and an algorithmic task. 
Validating tasks is difficult, because the real-world problem might not yet be fully understood (unclear reference), the problem might not have yet occurred in the real world (hypothetical reference), or the change of the problem is difficult to predict (future reference)~\citep{pidd2003}. 
Problem structuring methods~\citep{mingers2004problem} as well as validation and testing approaches from operations research~\citep{pidd2003}, requirements engineering~\citep{bourque2014guide} or design studies~\citep{munzner2014visualization} can be used for assuring validity.

\subsection{Methods for Generating Knowledge of Algorithm Designs}
Algorithm designs are neither arbitrary nor self-evident. 
We call those research methods that support the generation of engineering knowledge of algorithm designs \emph{design methods}. The outputs of these methods are essentially new algorithm designs. We distinguish deductive, inductive, abductive, and analogy engineering methods.

New algorithms can be developed by \emph{deduction}. In essence, deduction implies an operation of specialization. The starting points of deduction can be existing algorithmic tasks and designs as well as more general design principles, such as divide-and-conquer.

\begin{example}
\citet{ganapathi2022parallel} present divide-and-conquer for parallelizing classical iterative sorting algorithms, including a recursive algorithm that combines quicksort and bubble sort. \textit{``The aim is to sort the entire array $A[0\dots n-1]$. The function BubbleSort$(A[l\dots h]$) sorts the subarray $A[l\dots h]$. The initial invocation to the algorithm is BubbleSort($A[0\dots n-1]$). The function in turn calls the Partition function. The Partition function brings the smallest $n/2$ elements to the left half and the largest $n/2$ elements to the right half of array A. Once the array A is partitioned, then BubbleSort is recursively called onto the left and right halves in parallel to sort the two halves. After the two halves are sorted recursively, the entire array $A[0\dots n-1]$ will be sorted. When a subproblem reaches the base case, it is sorted using the standard iterative bubble sort logic.''} In this way, the authors deductively construct a new algorithm based on 1) the divide-and-conquer design principle also employed within quicksort and 2) the original bubble sort algorithm.
\end{example}

New algorithms can be developed by \emph{induction}. In general, induction means generalizing from a set of specific problem instances. According to \citet{manber1988using}, particularly the analogy to mathematical induction is a promising methodological reference for algorithm design. Mathematical induction is a method for proving that a statement $P(n)$ is true for all $n \in \mathbb{N}$. The proof by induction works by showing that 1) $P(n)$ holds for a base case such as $n = 0$ or $n = 1$ and 2) $P(n)$ holds for $n+1$. \citet{manber1988using} argues that this method can be transferred to algorithm design by solving an arbitrary instance of the problem at hand by assuming that the same problem has already been solved for a smaller size. 

\begin{example}
The application of inductive algorithm design can be illustrated using sorting algorithms~\cite{manber1988using}:
``\textit{[...] given a sequence of $n > 1$ numbers to sort (it is trivial to sort one number), we can assume that we already know how to sort $n - 1$ numbers. Then we can either sort the first $n - 1$ numbers and insert the $n$th number in its correct position (which leads to an algorithm called insertion sort), or start by putting the $n$th number in its final position and then sort the rest (which is called selection sort). We need only to address the operation on the $n$th number.}''  
The proof technique \emph{loop invariant} can be seen as a direct translation
from mathematical induction to proving the correctness of iterative algorithms
such as insertion sort, see \citet[Ch.~2]{cormen2009introduction}.
\end{example}

It is worth pointing out that induction can also be used as a general bottom-up approach. Given a problem, the idea is to start by first solving a specific instance of this problem. By extending the obtained solution, one might then be able to solve the original, larger problem.
\begin{example}
To illustrate this more general notion of the inductive approach, consider the algorithmic task of \textit{schema matching}. In essence, schema matching is the problem of generating correspondences between elements of two database schemas, e.g. for the purpose of data integration \cite{rahm2001survey}. Given two database schemas and a respective ground truth (i.e., the correct correspondences as identified by a human), the inductive approach is to analyze a few of the correspondences from the ground truth and then devise a possible algorithmic strategy to identify these correspondences automatically, e.g. via simple string matching. By step-wise increasing the set of considered instances, the strengths and weaknesses of the solution can be assessed and, if needed, a better solution can be developed.        
\end{example}

New algorithms can also be devised by \emph{abduction}~\citep{tomiyama2003abduction}. In essence, abduction is driven by anomalies~\citep{saetre2021generating}. In the context of algorithms, such anomalies might be related to specific circumstances in which an algorithm does not meet its expectations. The abductive process takes these anomalies as a starting point for detailed diagnosis, which provides hunches for generating plausible modifications and extensions. These can then be evaluated to which extent they meet expectations. 
\begin{example}
An example from process mining of abductive algorithm design is the $\alpha$-algorithm for generating Petri net models from a set of execution sequences, so-called event logs. The original $\alpha$-algorithm by~\citet{van2004workflow} is able to rediscover a Petri net from its set of execution sequences if this Petri net is structured and does not contain short loops (i.e., it does not produce execution sequences such as $aa$ or $abab$).  These limitations are addressed by later extensions. \citet{de2004process} propose the $\alpha$+algorithm. It offers a solution to the short-loop problem by preprocessing the input sequences. \citet{wen2007mining} work on the limitation that the original algorithm assumes that sequences are generated from a free-choice Petri net. The authors present the $\alpha$++algorithm, which is able to deal with non-local dependencies thanks to a modification of behavioural relations used by the algorithm.
\end{example}

New algorithms can also be developed by means of \emph{analogy} or \emph{metaphor}~\citep{sorensen2015metaheuristics}. This is the case when specific observations are generalized and detached from its context. Many algorithms stem from such generalizations. Most prominent are metaheuristics and optimization algorithms inspired by the phenomena discussed in the natural sciences such as simulated annealing~\citep{kirkpatrick1983optimization}, particle swarm optimization~\citep{kennedy1995particle}, or backpropagation for artificial neural networks~\citep{werbos1994roots}. Note that here we look at how these original algorithms have been formulated for the first time and not how they are applied to specific problems later. That latter case would be a deductive usage of existing algorithms.

\begin{example}
\textit{Genetic algorithms} are inspired by the process of natural selection \cite{forrest1996genetic}. They build on the concepts of mutation, crossover, and selection to iteratively generate high-quality solutions to optimization and search problems. Among others, genetic algorithms have been used for image processing, scheduling problems, and parameter optimization in machine learning. 
\end{example}

Some comments are warranted. First, the inspiration base for new algorithms is not restricted to the body of knowledge of algorithm engineering and other sciences. \citet{feyerabend1980against} provides arguments why any sort of inspiration, even if factually wrong, can be fertile. 
\newtext{\citet{brown2005many} have emphasized the value of artistic contributions that exhibit its merits in terms of subjective elegance or beauty.}
Algorithm engineering like any other act of human problem solving builds on creativity, for which dozens of mechanisms including and beyond the ones mentioned above exist. For an extensive list see~\citet[p.295]{simonton2022serendipity}. Also serendipity is important to mention in this context as accidental discovery of things not sought for. It is estimated that one quarter of scientific discoveries are serendipitous~\citep{thagard2012creative}.
Second, even if sought systematically, solutions in form of algorithms do not directly derive from algorithmic tasks. Design studies have shown that an understanding of the problem leads to solution ideas, which in turn provide a better problem understanding, and so forth. \citet{dorst2001creativity} refer to this as a co-evolution of problem and solution space. Similar observations have been made for system development by~\citet{guindon1990designing} and innovation management by~\citet{von2016crossroads}. These findings describe design as an opportunistic thought process. This means that creativity and serendipity leave a white space for algorithm engineering that has not yet been fully conquered by systematic methods.

No matter which engineering method is used, a key concern for developing engineering know\-ledge of algorithms in this context is \emph{design validity}. Adapting the definition by \citet{larsen2020validity}, we define this validity as the degree to which the internal structure of an algorithm is consistent, transparent, and explainable. The importance of explanation has been recently acknowledged by research on explainable artificial intelligence and decision making~\citep{saeed2023explainable}. Already \citet{naur1985programming} stated that a computer program can only be modified systematically if the developer has a theory about how the program works. In this way, design validity enables reproduction, critical assessment, and reuse in an incremental scientific process.

\subsection{Methods for Generating Formal Knowledge about Tasks and Designs}
Properties of algorithmic tasks and the outcomes of specific algorithms can be investigated using formal analysis methods and algorithm theory. In essence, such analysis is a branch of theoretical computer science. For that reason, analytical arguments are typically presented as theorems, for which their logical validity is demonstrated by corresponding proofs.

Analytical statements can be made about the complexity of algorithmic tasks and about the outcomes of specific algorithms. 
We associate outcomes with the efficiency of its computation and the effectiveness of the output. 
The \emph{efficiency} of computation is typically assessed by means of asymptotic analysis. To this end, asymptotic formulas are used to describe the magnitude of computational steps and memory space required. These are expressed according to~\citet{landau1909handbuch} as $\mathcal{O}(\cdot)$, $\Omega(\cdot)$, and $\Theta(\cdot)$.
In essence, $\mathcal{O}(\cdot)$ indicates the upper bound of the growth rate of a function that describes the efficiency, while $\Omega(\cdot)$ defines a lower bound, and $\Theta(\cdot)$ a bound above and below some strictly positive function. 
The symbols $o(\cdot)$ and $\omega(\cdot)$, in turn, refer to upper and lower
bounds, respectively, that are asymptotically not tight~\cite[Ch.~3.1]{cormen2009introduction}.
Identifying such efficiency bounds is a key research objective of algorithm theory. 

\begin{example}
\citet{bodlaender2015deterministic} investigate algorithms for connectivity tasks\footnote{We adopt terminology of our framework here. Algorithm theory typically uses the term \emph{problem} for what we call here \emph{algorithmic task}.} such as traveling salesperson, the number of Hamiltonian cycles, and the number of Steiner trees of a given graph $G=(V,E)$. They identify time complexity bounds for these problems that depend on the pathwidth and the treewidth of the input graphs. More specifically, \citet{bodlaender2015deterministic} formulate theorems and present corresponding proofs that algorithms with run-time complexity exist that challenge previously known bounds. Their bounds for
these NP-hard problems are exponential in the treewidth $\mathrm{tw}$ (or in the pathwidth), but
polynomial in the input size, i.e., $c^{\mathrm{tw}} \cdot |V|^{\mathcal{O}(1)}$ for 
constants $c$ that differ depending on the actual algorithmic task.
\end{example}

The \emph{effectiveness} of output is typically assessed in terms of correctness and completeness. Note that for certain notions of output effectiveness like approximating human judgment or human recognition of visual objects, the criteria of correctness and completeness do not objectively apply, such that effectiveness can be better understood as a notion of accuracy or solution quality. Many algorithms allow to trade off accuracy with efficiency.

\begin{example}
\citet{fan2010graph} present an algorithm that performs graph pattern matching in cubic time if graph patterns are restricted to specific connectivity properties. They prove the correctness by showing that it terminates, that the result is indeed a match, and that this match is maximum. Also a proof for the cubic time complexity is provided.
\end{example}

Formal knowledge about algorithm designs is established by the help of mathematical proofs. A mathematical conjecture for which a corresponding proof was found is called a \emph{theorem}~\citep{velleman2006prove}. Proofs make use of logical equivalences and inference rules for propositional or higher-order logic and for arithmetic.

\begin{example}
\citet[p.165-170]{gauch2003scientific} summarizes useful equivalences and inference rules for propositional logic. Equivalences include double negation, conditional exchange, commutativity, distribution, associativity, contraposition, DeMorgan's rules, exportation, redundancy, and biconditional. \citet[p.169-170]{gauch2003scientific} also states useful inference rules including modus ponens, modus tollens, simplification, conjunction, disjunctive syllogism, hypothetical syllogism, addition, and constructive dilemma. These rules build on the three axioms of identity, excluded middle, and noncontradiction.
\end{example}

Different methods for constructing proofs have been described including the construction of truth tables and deduction~\citep{gauch2003scientific}. Methodological support is provided by \citet{velleman2006prove}. He distinguishes different types of \emph{goals} that have to be proved and corresponding proof strategies.
\begin{enumerate}
    \item To prove $P \rightarrow Q$, assume $P$ is true and then prove $Q$.
    \item To prove $P \rightarrow Q$, assume $Q$ is false and then prove $P$ is false.
    \item To prove $\lnot P$, try to reformulate the goal and use another strategy.
    \item To prove $\lnot P$, assume $P$ is true and reach a contradiction.
    \item To prove $\forall x~P(x)$, assume that $x$ is arbitrary and prove $P(x)$.
    \item To prove $\exists x~P(x)$, choose specific $x$ and prove $P(x)$.
    \item To prove $P\land Q$, prove $P$ and $Q$ separately.
    \item To prove $P \leftrightarrow Q$, prove $P \rightarrow Q$ and $Q \rightarrow P$ separately.
    \item To prove $P \lor Q$, distinguish cases and prove either $P$ or $Q$.
    \item To prove $P \lor Q$, assume $P$ is false and prove $Q$.
    \item To prove $\exists !x P(x)$ (uniqueness), prove that $\forall y,z: P(y) \land P(z) \rightarrow y=z$.
    \item To prove $\exists !x P(x)$ (uniqueness), prove that $\exists x: P(x) \land P(y) \rightarrow x=y$.
\end{enumerate}

\begin{example}
\citet{lokshtanov2018known} investigate lower bounds of run-time efficiency of algorithms operating on graphs with bounded treewidth. They present several theorems and lemmata with corresponding proofs. Their Theorem 1 has an implication as a goal and is proved using Velleman's Strategy (1) by construction. Their Lemma 11 includes three conjectures. For the first and the second, Strategy (9) is applied distinguishing two cases. For the third, Strategy (2) is used for constructing a contradiction.
Lemma 15 is proved by help of Strategy (6). An arbitrary assignment is chosen for which the conclusion is proved. Many figures showing example graphs illustrate the idea of the proofs and concepts.
\end{example}

The key concern of proofs for theorems is logical validity. 
\newtext{We define \emph{logical validity} based on Aristotle as the degree to which the syllogisms used in a proof preserve truth~\citep[p.374]{durand2008truth}.}
Crafting proofs is often a manual exercise performed by researchers. Tools can be used to support this exercise. Techniques from model checking and verification have been implemented in \emph{theorem provers} such as the ones by
\citet{paulson1994isabelle,schulz2002brainiac,detlefs2005simplify}. 
\newtext{Still, as \citet{lakatos2015proofs} emphasizes, the eventual confidence in the logical validity of a proof is hardly a uni-directional deductive exercise, but rather a social process within the research community.}

\subsection{Methods for Generating Empirical Knowledge about Algorithm Designs}
Empirical knowledge about algorithm designs refers to algorithmic tasks, designs, and their relationship. Such empirical knowledge can be methodologically established by \emph{developing hypotheses} and testing them. New hypotheses can be derived inductively or deductively~\cite{recker2012scientific}. The importance of theories as a knowledge context~\citep{DBLP:books/sp/Wieringa14} has been stressed in various empirical branches of computer science such as software engineering~\cite{johnson2012s,wohlin2015general,DBLP:journals/tse/Ralph19} and computer visualization~\cite{sedlmair2012design}.

\subsubsection{Develop Hypotheses}
The \emph{inductive development of hypotheses} takes World 1 and its entities as a starting point. This includes the algorithmic problem, available implementations and data, as well as prior outcomes. \citet{popper1977worlds} states that \textit{``in order to learn more about World 1, [the researcher] must theorize.''} Such \emph{theorizing} is subjective to thoughts and observations about World 1 by the researcher in World 2 yielding hypotheses as World 3 entities.

The \emph{deductive development of hypotheses} takes World 3 and its different knowledge entities as a starting point. 
A reference for such deduction can be formal knowledge about designs and tasks, which may have been formally proved in earlier publications. Also empirical knowledge about designs and tasks published in prior research can inform the formulation of new hypotheses. Such knowledge can also stem from other disciplines. 

\begin{example}
\citet{sedgewick1978implementing} discusses the implementation of quicksort. Prior \emph{formal} knowledge states that the average-case run-time performance increases with input size $n$ according to $\mathcal{O}(n \log n)$. Sedgewick posits that the run time of a practical quicksort implementation can be improved -- compared to the standard algorithm design found in textbooks. He relies on prior \emph{empirical} (and to some extent also \emph{formal}) knowledge on issues concerning recursion depth, small subarrays, worst-case input distributions, and different partitions.
\end{example}

\begin{example}
The graph drawing discipline develops algorithms that produce effective and aesthetic layouts for graphs. \citet{ware2002cognitive} propose a set of metrics of a graph layout including the number of crossings, path bendiness, or shortest path length. Their experimental study shows that these metrics are correlated with cognitive effectiveness of the graph layout for human participants. The justification such metrics builds on cognitive theories~\citep{malinova2021cognitive}. Based on metrics of graph aesthetics, algorithms have been compared and further developed~\citep{gibson2013survey}.
\end{example}

It is important to note that hypotheses are not always articulated in an explicit manner. This, however, does not mean that no hypotheses exist. If, for instance, the performance of a novel design is compared against the state-of-the-art design without stating a clear hypothesis, there still exists the implicit hypothesis that the novel design is better with respect to a relevant performance dimension.   

The key concern for hypothesis and theorem formulation is \emph{justification validity}. 
It refers to \emph{convincingly} a hypothesis or theorem is supported by a deductive argument~\citep{lannin2005generalization}. 
Depending on the type of hypothesis, this might be a matter of deduction from prior knowledge and, if applicable, its underlying theoretical justification. 
In certain settings, the backing of a theoretical justification is essential for establishing an explanation for the expected effect as stated by the hypothesis. It should again be noted that not all hypotheses in algorithm engineering are or can be justified by theory. Often, nonetheless, there is a rationale why a certain association, e.g. between a design and performance, should be present.

\subsubsection{Derive Research Design}
The term \emph{research design} refers to the plan used to examine a research question of interest~\cite{marczyk2005essentials}. Various types of design can be distinguished according to how explicitly they define and justify hypotheses and control for alternative explanations. An \emph{exploratory design} imposes no control and does not rely on justified hypotheses. It is appropriate in areas where algorithmic tasks and algorithm designs are not yet well understood. 

A \emph{correlational design} identifies hypotheses and corresponding measures. It does not control for alternative explanations and does not make use of explicit treatments. It can provide evidence for empirical connections, but does not offer causal insights. Such connections, for instance between the different data characteristics and performance, can be assessed using correlation and regression. 
Correlational studies have been used in empirical software engineering in the early 1980s, e.g.~\citet{basili1983metric}, but were later largely replaced by experiments for better control of confounding factors.

An \emph{experimental design} manipulates one or more experimental factors with two or more levels in a controlled way. 
Often, the algorithm design is considered as this factor with several alternative algorithms representing the different levels. Also properties of the input data can be varied. 
The effects of the factor can be evaluated using statistical tests of mean comparison such as ANOVA or non-parametric alternatives~\cite{demvsar2006statistical}.

\begin{example}
In the field of process mining,~\citet{augusto2018automated} conducted an exploratory study comparing the different algorithms for generating process models from event logs. These models are evaluated using accuracy and complexity measures, showing relative strengths and weaknesses of different algorithms. A follow-up correlational study by~\citet{augusto2022connection} investigates how properties of the input data are connected with properties of the generated process models by estimating correlation coefficients and regression functions. 
Experimental studies in this area are scarce. Experiments require the explicit manipulation of, for instance, event log properties by generating artificial event log data with controlled properties. An example of such an experimental study is \citet{janssenswillen2019towards} who investigate the relation between process discovery algorithms and the fitness-precision trade-off in an experimental setting.
\end{example}

A central concern of the research design is \emph{internal validity}. A research design is internally valid if its manipulation is causally responsible for an observed effect~\citep[p.102,106]{wohlin2012experimentation}. Various factors that relate to the real-world problem, the algorithm itself, and the execution environment can affect validity~\citep{barr1995designing}.
Exploratory and correlational designs are weak in terms of internal validity. Experimental designs build on randomization and blocking to eliminate the effect of potentially confounding factors~\cite{wohlin2012experimentation}. Randomization is a means to statistically eliminate the effect of confounding factors by randomly assigning units to treatment groups. Blocking eliminates the effect of confounding factors by keeping them at a constant level. For instance, consider a new algorithm is claimed to perform better than a recent alternative. Then, both can be run repeatedly on the same machine to avoid confounding effects of memory usage by randomization. Furthermore, both can be implemented in the same programming language by the same developer to block these factors~\cite{kriegel2017black}.

\subsubsection{Build Implementation for Instrumentation}
The \emph{instrumentation} of the research design instantiates the research design by running an implementation of the algorithm with evaluation data as input to produce performance measurements of the computational process and of the generated output.
An \emph{implementation} of the algorithm design is used for the instrumentation.
Any implementation uses a specific programming language on a specific operating system running on a specific microprocessor family, which all affect the performance of the algorithm.

\begin{example}
\citet{kriegel2017black} demonstrate that different implementations of the same algorithm design can vary in performance by orders of magnitude. Run time of different DBScan implementations do not only differ by four orders of magnitude. Even using the same implementation languages, but in different versions of the same frameworks like ELKI and WEKA yielded substantial performance differences.
\end{example}

Several publications give practical advice and clarify which matters to consider for implementing the infrastructure for instrumentation. Various practical concerns are relevant such as software testing, handling errors, tracking provenance, and creating packages, as much as matters of team work and appropriate use of version control and command-line tools~\citep{irving2021research}.

\begin{example}
\citet{DBLP:journals/algorithms/AngrimanGLMNPT19} present guidelines on how to build the experimental pipeline addressing the following concerns. Principles of software testing such as unit tests should be considered. Implementation code should be managed using version control systems and shared as open source on platforms like Github. In general, experiments should be fully automated using scripts or tools such as SimexPal in order to provide reproducibility. Output files should be separated into experimental results, metadata, and supplementary data, in both a human readable, but also machine processable way.
\end{example}

A key concern here is~\emph{implementation validity} (or instantiation validity~\cite{lukyanenko2015guidelines,lukyanenko2020design}).
One threat for implementation validity is that the design generally leaves space for alternative implementation options. In this way, implementation decisions can turn out to be confounding factors~\cite{lukyanenko2015guidelines}. 
Another threat is a potentially unfaithful implementation. Testing cannot fully assure the conformance with the design specification. Optimization, debugging and publishing code help to establish the desired confidence~\cite{kriegel2017black}.
Formal verification with corresponding tools, if applicable, can assure that an implementation meets its design specification \cite{4544862}.

\subsubsection{Choose Evaluation Data for Instrumentation}
Algorithms have to be run on input such that we can empirically investigate their performance. This input for the instrumentation is called \emph{evaluation data}. 
Different types of evaluation data can be distinguished along two dimensions: whether the data is publicly available or not and whether it stems from real-world or artificially generated problem instances.

In many research communities, \textit{benchmark data sets} are made publicly available in order to encourage the performance evaluation of competing algorithms. The utilization of benchmark data sets affects the validity of research findings for several reasons. On the one hand, using benchmark data acts as a blocking mechanism and eliminates hidden confounding effects that could arise from using disparate data sets. On the other hand, the prolonged use of the same benchmark data could seduce researchers to tailor their designs to the characteristics of a specific data set. This, in turn, increases the risk of overfitting the specific benchmark data and hurting the \emph{external validity} of the research findings~\cite{sim2003using,tichy1998should}.

Benchmark data sets are available for various algorithmic tasks, such as part-of-speech tagging \cite{paul1992design}, image recognition \cite{russakovsky2015imagenet}, ontology matching \cite{algergawy2019results}, process mining \cite{DBLP:conf/bpm/DongenWFW13}, vehicle routing problems \cite{DEFRYN2016400}, network analysis \cite{DBLP:conf/www/Kunegis13,DBLP:journals/tist/LeskovecS16}, and various combinatorial optimization problems such as graph partitioning and graph clustering \cite{DBLP:reference/snam/BaderKM00W18}. Benchmark data sets not only facilitate comparative evaluation. They also play an important role for exploring how specific features of realistic input data can be exploited~\citep{sanders2009algorithm}. 

\begin{example}
Different versions of the ImageNet data set have been used for several classification challenges including the tasks of image classification, single-object localization, and object detection. \citet{russakovsky2015imagenet} provide an overview of the results for the years 2012-2014 with error rates and corresponding confidence intervals for many published algorithms.
\end{example}

Benchmarking data might not be available for new or specific algorithmic tasks or at least not at the required level of richness. In this case, the use of private data is appropriate, even if that data cannot be shared. 

\begin{example}
\citet{alsger2016validating} develop an algorithm to estimate origin–destination trails based on smart card fare data. The authors use a unique smart card fare data set obtained from the Australian public transport operator TransLink. This data set provides rich details including boarding and alighting times and locations for each passenger. Using this data set, the accuracy of the algorithm's capability to estimate origin–destination trails is evaluated.
\end{example}

Data sets, no matter if public or private, can be sampled from a real-world problem instance or artificially generated.
The TransLink data set is also an example of a \emph{real-world} data set. Real-world data sets are valuable for capturing the variability and complexity of a real-world problem, or at least the relevant and realistic range of to-be-expected input~\cite{sanders2009algorithm}.
\emph{Artificial} (computer-simulated or synthetic) data sets \cite{santner2003design} can be used once relevant input data features are understood. Simulation affords the systematic variation of typical input ranges, which is instrumental for generating knowledge into how input data features translate into algorithmic performance variations. Simulated data is also useful when empirical data does not exist, is insufficient, imprecise~\cite{hofmann2013ontologies}, or too large to store or share~\cite{DBLP:journals/corr/abs-2003-00736}.

\begin{example}
Feature selection algorithms identify relevant features in high-dimensional data. \citet{bolon2013review} review feature selection algorithms using artificial data for two reasons: \textit{``1. Controlled experiments can be developed by systematically varying chosen experimental
conditions, like adding more irrelevant features or noise in the input. This fact facilitates
to draw more useful conclusions and to test the strengths and weaknesses of the existing
algorithms.
2. The main advantage of artificial scenarios is the knowledge of the set of optimal features
that must be selected; thus, the degree of closeness to any of these solutions can be
assessed in a confident way.''}
\end{example}

The choice for real-world or artificially generated data sets has implications for external validity.
\newtext{\emph{External validity} is the degree of how well we can generalize results across populations of interest~\cite[p.467]{cook2002experimental}, i.e. data sets of interest in our case.}
One aspect of external validity relates to \emph{statistical generalizability}. This is the question whether the results obtained from a study can be generalized to the larger population from which the data was extracted~\cite{winer_experimentation_1999}. Ideally, the data should be a random sample from the population of interest, as it simplifies the task of achieving statistical generalizability. This approach is quite straightforward when dealing with artificial data. However, when dealing with real-world data, random sampling is often not feasible. In such cases, \citet{cook2002experimental} recommend employing purposive sampling to increase heterogeneity among the instances. In line with this argument,~\citet{kriegel2017black} recommends to vary features related to the size and complexity of the input data.

\begin{example}
\newtext{\citet{van2019sok} identify frequent benchmarking flaws in systems security evaluations, such as cherry-picking workloads, using unrealistic threat models, or omitting baseline comparisons. These issues pose a threat primarily to external validity, as the evaluation results often fail to generalize beyond the constrained conditions under which they were obtained. Although the benchmarks may be internally consistent, they do not reliably represent real-world environments and, thus, limit the applicability of the conclusions drawn.}
\end{example}

Another aspect of external validity pertains to \emph{realism}. This is the question to which extent the findings can be applied to a more natural or real-world setting. Real-world data, often referred to as "\textit{organic data}," are generally assumed to possess such realism. Nevertheless, as long as there are unknown discrepancies between the data generation processes across different populations, potential threats to external validity also exist for real-world data~\cite{xu2019validity}.
Achieving realism poses a greater challenge when working with artificial data. Experimental settings using artificial data often tend to simplify reality in order to unveil causal patterns, which may  limit their applicability to more natural settings~\cite{DBLP:journals/scp/WieringaD15}. The primary challenge when dealing with artificial data in this regard lies in the endeavor to study features of real-world problems and ensure that these features are adequately reflected in the artificial data~\cite{lhermitte2011comparison}.

\begin{example}
\citet{bolon2013review} conduct an extensive study using artificial data sets. 
In order to strengthen the external validity of their analysis, they extend their study to additionally include two real-world data sets. In this way, they demonstrate that conclusions of their study with artificial data sets also extend to real scenarios.
\end{example}

\subsubsection{Measurement and Instrumentation}
Hypotheses on algorithms have to be operationalized to \emph{measures of performance}. 
\newtext{\citet{larson2025} use the term \emph{criterion}.}
These measures 
capture the efficiency of the algorithm execution and the effectiveness of the generated output. The measurement of \emph{efficiency} is classically based on the number of steps an algorithm performs until termination or run-time duration~\citep{knuth1974computer}. The \emph{effectiveness} of the output can be measured in various ways based on the distance between the desired and actually generated output or an assessment of the usefulness of the output.

\begin{example}
Graph drawing algorithms address the real-world problem of visualizing an abstract graph in an efficient and effective manner. Efficiency is often based on the run time and measured using, for instance, the CPU time in seconds~\citep{hachul2005experimental}. Effectiveness is often assessed according to how well the algorithm's output can assist humans solving real-world problems, for instance, as measured by the speed and accuracy of human task performance~\citep{purchase1998performance}. 
\end{example}

A key concern of measurement is \emph{construct validity}. In essence, it points to the question  whether a measure of a construct sufficiently measures the intended property \cite{o1998empirical}. More specifically, the measures have to be a valid and reliable operationalization of the intended concept~\cite{DBLP:books/daglib/0084392}. For run-time \emph{efficiency measurement}, \citet{kriegel2017black} call for caution with wall-clock time, since it can be dominated by implementation details.
\emph{Effectiveness measurement} comes with more challenges and relates to properties of what is called gold standard, ground truth, reference data, or test data~\citep{kondermann2013ground,zendel2017good,TOFT200519}. A gold standard is not the data that is processed by an algorithm, but the reference against which the generated output is compared.

If an objectively correct or optimal gold standard exists, the effectiveness of the solution can be assessed by the help of simple distance measures or measures such as precision, recall, mean square error, or area under the curve~\cite{hossin2015review,saito2015precision}. A gold standard can be constructed by means of highly accurate devices, by simulating, or by generating synthetic data~\citep{kondermann2013ground}.   
If a gold standard is obtained by annotation of human judgement, the same measures can be applied, but have to be interpreted differently. Unknown human errors or biases likely exist depending on the type of task~\citep{kondermann2013ground}. Research on ImageNet has, initially as a by-product, yielded extensive insights into sophisticated large-scale annotation procedures using crowdsourcing~\citep{russakovsky2015imagenet}.
If a gold standard is absent, human judgement can be used to assess the usefulness of the output. This assessment can be based on psychometric measurement using questionnaire items, for instance, of the technology acceptance model~\citep{venkatesh2000theoretical}. Psychometric measurement items can also be newly developed using guidelines such as the ones proposed by~\citet{petter2007specifying} and summarized by \citet[p.99]{recker2012scientific}.

\begin{example}
Process drift detection is the analytical task of automatically identifying those points in time where the behaviour of a process changes. Algorithms for such drift detection take event sequences as input. \citet{maaradji2017detecting} evaluate their algorithm in terms of wall-clock time efficiency and effectiveness in terms of how precision and recall develop relative to time window size. \citet{DBLP:journals/corr/abs-1907-06386} evaluate their drift detection algorithm for efficiency using wall-clock time and effectiveness based on user assessment of usefulness of the generated output. This assessment uses items of the technology acceptance model~\citep{venkatesh2000theoretical}.
\end{example}

Research designs involving human participants have been critically appraised from an ethical standpoint. Awareness of ethical issues sprang from experiments that could cause medical and psychological harm for participants~\citep{rutstein1969ethical,slater2006virtual}. Classical designs such as the Milgram experiments on obedience are distant to typical algorithm engineering research; however, ethical concerns are relevant, for instance, in studies where algorithmic output is evaluated using man-made gold standards.

\begin{example}
\citet{miceli2020between} conduct a participatory study on data annotation of images by humans in two annotation companies, one in Buenos Aires, Argentina, and one in Sofia, Bulgaria. Their findings highlight intertwined issues with biases, misunderstanding, and unawareness on the one hand and personal vulnerability of the individual workers on the other hand. As a consequence, \citet{kazimzade2020biased} recommend the development of ethical standards for data annotation in terms of transparency, education of annotators, and corresponding regulation.
\end{example}

\subsubsection{Draw Conclusions}
The \emph{comparison} of the expected with the observed results leads either to a rejection or a failure to reject the null hypothesis. In case of an \emph{unexpected} failure to reject the null hypothesis, it is challenging to identify the cause of the deviation. 
The reason for failure can be a mismatch between real-world problem and implementation in World 1, human misinterpretations in World 2, or failures across worlds between algorithm task and real-world problem, algorithm design and task, or algorithm implementation and design~\citep{staples2015-methodology}. Methods for root cause analysis~\citep{doggett2005root} can be used to single out what did work and what not.
Upon this basis, there are two ways of responding to such unexpected results: a revision of the algorithm task, design, or implementation (\emph{design revision}) or a revision of the hypotheses, research design, instrumentation, or measurement (\emph{knowledge revision})~\citep{staples2015-methodology}. In this way, scientific knowledge creation manifests itself as an iterative process that is directed towards aligning the algorithmic task and design, and any corresponding knowledge.

\begin{example}
\citet{pittke2015automatic} present algorithms to detect homonyms and synonyms in text labels of process model collections. Their evaluation analyzes efficiency in terms of processing time and effectiveness using precision and recall. Furthermore, they present examples to illustrate the capabilities of their approach. The examples highlight that resolving one case of a homonym or synonym potentially creates issues with other terms of the process model collection. This unexpected insight points to the opportunity of a design revision in future research.
\end{example}

In order to gain a profound understanding of the mechanisms that explain a certain result, it has been recommended to use a combination of macro- and micro-benchmarks. 

\begin{example}
\newtext{\citet{traeger2008nine} describe various considerations for a careful evaluation design for file system and storage benchmarking. Specifically, the authors propose using at least one macro-benchmark, capturing a full application workload, with a diverse set of targeted micro-benchmarks, designed to analyze specific file system behaviors, such as small writes or metadata operations. This design supports internal validity, as the micro-benchmarks isolate and evaluate individual performance aspects with minimal confounding. At the same time, the inclusion of the macro-benchmark contributes to external validity by ensuring that overall system behavior is assessed under realistic, application-driven conditions. The complementary use of both types of benchmarks allows for fine-grained analysis while maintaining relevance to real-world workloads.}
\end{example}

Also the \emph{expected} rejection of the null hypothesis requires some further reflection. 
The key concern here is \emph{conclusion validity}, i.e.\ to which degree the 
results can reasonably be regarded as revealing the hypothesized connection~\cite{cook1979design,garcia2012statistical}. Conclusion validity strongly emphasizes statistical analysis, but also covers qualitative considerations \cite{cozby2007methods}. Conclusions can be drawn from statistical tests when their assumptions hold and the required significance levels are obtained. Additional qualitative analysis of outliers or data points that exhibit anomalies can help to further support  conclusions~\cite{hernandez2017evaluation} or to investigate reasons of failure. Often, reservations on conclusions from empirical research are reported as \emph{limitations} or \emph{threats to validity}.

\begin{example}
\citet{teymourian2020fast} present a clustering algorithm for the modularization of large-scale software systems. The authors conduct extensive experiments to show that their solution can compete with existing algorithms in terms of both modularization quality as well as run time. At the end of the paper, the authors have included a dedicated section entitled "\textit{Threats to Validity}". They use this section to mainly discuss concerns of external and internal validity. Among others, they discuss that the selected application might not be representative of software systems in general and that employing different evaluation metrics may lead to different results. Besides external and internal validity, the authors also discuss additional validity concerns, such as conclusion validity. For instance, they point out that the employed ground truth might not actually match the ground truth created by humans.  
\end{example}

Researchers have to take a step back and stick to the facts when drawing conclusions.
Results, no matter if expected or unexpected, may be subject to errors or fraudulent manipulation. \newtext{Imperative for understanding is that all results can be explained~\citep{small1997does}.} \citet{gauch2003scientific} states that 
``\emph{most fallacious reasoning in science results from accidental
blunders. Unexamined presuppositions, bad data, and invalid logic
lead to wrong conclusions, despite the best of intentions. To be honest, however,
it must be admitted that scientists are only human, so occasionally errors
result not from failure of competence but from failure of will. $[ \dots]$ Sometimes reason
is usurped by desire, so the goal becomes not to embrace reality, but to evade
reality. Logic is enlisted in this dirty, insincere business of rationalizing.}''

\emph{Reproducibility} is therefore a key concern for establishing transparency and trust in scientific results. In essence, reproducibility requires that the algorithm can be implemented and evaluated independently by a different team~\citep{DBLP:journals/algorithms/AngrimanGLMNPT19}. Weaker notions include replicability (a different research team conducts a study using the same source code and experimental setup) and repeatability (the same research team conducts a study using the same source code and experimental setup)~\citep{plesser2018reproducibility}.
To meet reproducibility as a requirement, evaluation data and implementations should be made available and citable~\citep{stodden2016enhancing}. Different initiatives, such as making research data and research software meet the principles of findability, accessibility, interoperability, and reusability~(FAIR), address this concern~\citep{wilkinson2016fair,barker2022introducing}, not only to reproduce results, but also to reuse research material to help gaining new insights.

\begin{example}
The Proceedings of the VLDB Endowment encourages authors to share their work in a reproducible way. Corresponding submissions are invited to include the following material:\footnote{\url{https://vldb.org/pvldb/reproducibility/}, accessed 7 August 2023.} a prototype system implementation, either the process to generate the input data or the actual data itself, the executable code as instrumentation to produce the raw result data, and the scripts to transform the raw data into the graphs included in the paper.
The paper by~\citet{tziavelis2020optimal} on optimal algorithms for ranked enumeration of answers to full conjunctive queries is one of the accepted VLDB papers that has been evaluated to meet these reproducibility criteria. 
\end{example}

\section{Implications for Algorithm Engineering}
\label{sec:discussion}
This section discusses the implications of our framework for algorithm engineering. The framework highlights that research papers in this area require ontological clarity, epistemological precision, and methodological validity. 
These three pillars provide the foundation for a scientific contribution. 
In this context, a \textit{contribution} is an extension or an improvement of the body of knowledge regarding \textit{knowledge of} tasks and design, \textit{knowledge about} tasks and design, and \textit{knowledge how} to conduct research in terms of methodology. The accumulation of scientific contributions in the body of knowledge represents scientific progress.

In the following, we discuss which questions help capturing to which extent a contribution has ontological clarity, epistemological precision, and methodological validity.

\subsection{Ontological Clarity}
In essence, a contribution has \textit{ontological clarity} if it is clear to which algorithmic task and algorithm design it relates, as much as corresponding real-world problem, implementations, and results. Table~\ref{tab:onto} summarizes questions that authors should consider answering in their papers with respect to \textit{tasks}. The audience of the paper must be able to understand what the exact task is and to which real-world problems it relates. Some tasks are canonical and well understood in specific research areas. The more specific and newly identified a particular task is, the more detailed elaboration on its characteristics is required, for instance using models and examples. Clarity in this matter provides a basis for judging the appropriateness of assumptions in input data and algorithmic processing, as much as on the desired output and performance requirements. If the task cannot be fully specified or solved algorithmically, it should be clarified how users can interactively direct the computation and work with the output.

Table~\ref{tab:onto} also summarizes questions that authors should consider in their papers with respect to \textit{designs}. The audience of the paper must be able to understand how the algorithm works and how this addresses the requirements imposed by the task. Clarifying the design principles upon which the design rests, how design decisions were taken, and how parameter values were set, helps readers to judge plausibility.
If the contribution relates to an implementation and corresponding experiments, relevant additional implementation decisions should be explained and made transparent. 

\begin{table}[bt]
    \footnotesize
    \centering
    \begin{tabular}{|p{.95\textwidth}|}
        \hline
        \textbf{Ontological Clarity of Tasks}\\
         \hline
            What is the task? \\
            How does the task relate to real-world problems?\\
            Which assumptions are made regarding input data?\\
            Which assumptions are made regarding algorithmic processing?\\
            What is the desired output?\\
            Which performance requirements are relevant?\\
            To which extent can the task be fully specified?\\
            How are users involved if the task is fuzzy?\\
        \hline
        \textbf{Ontological Clarity of Designs}\\
         \hline
            How does the algorithm work?\\
            Upon which design principles is the algorithm designed?\\
            Which design decisions were made based on which assumptions?\\
            Based on which considerations have parameter values been set?\\
            If applicable, which implementation decisions have been made?\\
            Which results does the implementation provide?\\
        \hline
    \end{tabular}
    \caption{Questions on Ontological Clarity}
    \label{tab:onto}
     \vspace*{-0.75cm}
\end{table}

\subsection{Epistemological Precision}
A contribution has \textit{epistemological precision} if it precisely describes what is not known, how that lack of knowledge constitutes a research problem, and how it extends the body of knowledge in this respect. Table~\ref{tab:epis} summarizes related questions on \textit{tasks}. 
The task can be described in different ways. The anticipated contribution defines the degree of precision to which the task has to be specified. If possible, descriptions from prior research should be used or appropriately adapted. Prior knowledge of a task defines the backdrop against which corresponding knowledge about the task can be identified. 
Understanding the boundaries of the body of knowledge requires a comprehensive review of the relevant literature, including formal and empirical contributions. Mind that not only publications on the exact task are relevant, but also more general classes of tasks and related tasks inform our understanding.
The systematic review of the literature is the basis for summarizing what is known and the new or improved knowledge that a paper provides.

Table~\ref{tab:epis} also summarizes related questions on \textit{designs}. Similar concerns apply as for tasks. 
The anticipated contribution defines the degree of precision to which a design has to be specified. If possible, descriptions from prior research should be reused. Prior knowledge of a design defines the boundaries of relevant knowledge about it. Mind that this knowledge about designs is relative to the specific tasks under consideration.
A comprehensive review of the relevant literature on related designs is required. If possible, insights into performance should be complemented with analysis of sensitivity, uncertainty, and explanatory factors.

\begin{table}[tb]
\footnotesize
    \centering
    \begin{tabular}{|p{.95\textwidth}|}
        \hline
        \textbf{Epistemological Precision of Tasks}\\
         \hline
            Why is the task described formally, semi-formally, informally, or else?\\
            What prior research has been published on this task?\\
            What knowledge of this specific task has been published?\\
            Which new or improved knowledge of this task is presented?\\
            Which formal knowledge about this task including theorems has been published?\\
            Which new or improved theorems about this task are presented?\\
            Which empirical knowledge including hypotheses about this task has been published?\\
            Which new or improved hypotheses about this task are presented?\\
        \hline
        \textbf{Epistemological Precision of Designs}\\
         \hline
            What knowledge of a specific design has been published?\\
            Which new or improved knowledge of a design is presented?\\
            Why is the design described using pseudocode, flow charts, formal specification, or else?\\
            Which formal knowledge about a design including theorems has been published?\\
            Which new or improved theorems about a design are presented?\\
            Which empirical knowledge including hypotheses about a design has been published?\\
            Which new or improved hypotheses about a design are presented?\\
            Do theorems and hypotheses relate to performance, sensitivity, uncertainty, and explanatory knowledge with corresponding causal factors?\\
        \hline
    \end{tabular}
    \caption{Questions on Epistemological Precision}
    \label{tab:epis}
     \vspace*{-0.75cm}
\end{table}

\subsection{Methodological Validity}
A reader of a research paper needs to understand the boundaries of the knowledge that is being created.
A contribution has \textit{methodological validity} if the latest research methods and methodological concerns are appropriately considered. Overall, we distinguish nine concerns. Table~\ref{tab:method} summarizes corresponding questions. 

The central question regarding validity is which validity concerns have to be considered. First, this consideration depends upon the \textit{ontological focus of the contribution}. 
\begin{itemize}
    \item If a contribution relates to algorithmic tasks, then \textit{ecological validity} is relevant. 
    \item If a contribution presents an algorithm design, then \textit{design validity} is important. 
    \item If a contribution builds, among others, on an algorithm implementation, then \textit{implementation validity} has to be considered.  
\end{itemize}

Second, it is relevant to which \textit{type of knowledge} a contribution refers.
\begin{itemize}
    \item No matter if a contribution provides knowledge of or about tasks or design,  \textit{methodological transparency} is required. Based on which considerations and methods is this knowledge established?
    \item If a contribution offers new formal knowledge, then both \textit{justification validity} of establishing theorems and \textit{logical validity} of corresponding proofs has to be considered.
    \item If a contribution formulates new empirical knowledge, then \textit{justification validity} of the hypotheses and \textit{internal validity} of the research design have to be considered. Furthermore, the instrumentation of the research design requires the \textit{implementation validity} of the implementation and the \textit{external validity} of the evaluation data. \textit{Construct validity} is relevant for the results produced and \textit{conclusion validity} for drawing appropriate conclusions.
\end{itemize}

Third, there are different strategies to address validity concerns. The weakest consideration is to create transparency about limitations, for instance in a subsection on threats to validity. A better approach is to conduct additional analyses to investigate to which extent a certain aspect of validity could be affected by factors that were not addressed by the research design. If possible, an even better strategy is when validity concerns are anticipated and considered in the research design, for instance by help of manipulation checks, systematic sampling of evaluation data, or other means of quality control. All this helps to assure that the right conclusions can be drawn from the research at hand.

Fourth, it has to be noted that a contribution can never be perfect with respect to all validity concerns. In particular, the trade-off between internal and external validity has often been discussed in the literature. To a certain extent, this trade-off is inherited from the choice of a specific research design and research method~\citep{recker2012scientific}. One way to address this is to integrate multiple different research designs into a research strategy~\citep{roe2009internal}. Another perspective on imperfections is to judge them based on what is known on a particular research problem. If hardly any knowledge is available on an emerging research problem, the bar for methodological rigor might be lower than for well-researched problems. However, the bar is then still high to explicitly discuss threats to validity. 

In the end, it is important to emphasize that \emph{knowledge of} without \emph{knowledge about} is insufficient. The central knowledge concern of algorithm engineering is the satisfaction relationship between algorithmic tasks and algorithm designs. This implies that validity concerns are relevant for any paper that proposes a new algorithm.

\begin{table}[tb]
    \footnotesize
    \centering
    \begin{tabular}{|p{.95\textwidth}|}
        \hline
        \textbf{Methodological Validity of Tasks}\\
         \hline
            Which methods are used to generate knowledge of tasks?\\
            If there are relevant threats to \emph{ecological validity}, how have they been mitigated?\\
            To what extent can the study findings be generalized to real-world problems?\\
        \hline
        \textbf{Methodological Validity of Design}\\
         \hline
Which methods are used to generate knowledge of designs?\\
Why have deductive, inductive, abductive, analogy methods been used, or else?\\
How did the co-evolution of problem and design space unfold?\\
If there are relevant threats to \textit{design validity}, how have they been mitigated?\\
To which extent is the internal structure of an algorithm consistent, transparent and explainable?\\
       \hline
To which asymptotic function class is the efficiency related?\\
Which formal guarantees regarding correctness, completeness, and termination can be made?\\
If there are relevant concerns regarding \textit{logical validity}, how have they been addressed?\\
What is the trade-off between efficiency and effectiveness?\\
Which proof strategies are used to prove theorems and lemmata?\\
       \hline
Why are hypotheses developed deductively or inductively, or else?\\
If there are relevant threats to \textit{justification validity}, how have they been mitigated?\\
Do hypotheses have a theoretical justification or else?\\
        \hline
Why is an exploratory, correlational, or experimental research design used?\\
If there are relevant threats to \textit{internal validity}, how have they been mitigated?\\
What are potentially confounding factors and how are they controlled?\\
       \hline
Which implementation method was chosen and why?\\
How was the implementation tested?\\
If there are relevant threats to \textit{implementation validity}, how have they been mitigated?\\
        \hline
Which evaluation data is chosen and why?\\
Why is benchmark data or private data, real-world data or artificial used?\\
If there are relevant threats to \textit{external validity}, how have they been mitigated?\\
Are feature ranges understood and realistic?\\
How are relevant factors untangled and isolated?\\
How is overfitting avoided?\\
        \hline
How are appropriate measures selected or constructed?\\
If there are relevant threats to \textit{construct validity}, how have they been mitigated?\\
Are the measures valid and reliable operationalizations of the intended concept?\\
Is effectiveness measured by help of an objective gold standard, a gold standard obtained from human annotation, or human judgements in absence of a gold standard?\\
       \hline
How can unexpected results be explained?\\
What are arguments for a design revision or a knowledge revision?\\
If there are relevant threats to \textit{conclusion validity}, how have they been mitigated?\\
Do assumptions of statistical tests hold?\\
What can we learn from outliers or anomalous data points?\\
Are the results reproducible according to established standards?\\
Are data and research software shared upon FAIR principles?\\
       \hline
    \end{tabular}
    \caption{Questions on Methodological Validity}
    \label{tab:method}
     \vspace*{-0.75cm}
\end{table}

\subsection{Ethical Considerations along the Algorithm Engineering Framework}

Ethics plays a pivotal role in algorithm engineering since algorithms increasingly impact various facets of society. Recognizing this, many authors discuss numerous dimensions of ethics in the context of designing and using algorithms including trust, fairness, transparency, and bias mitigation \cite{barocas2023fairness,martin2019designing,spiekermann2022values}.
If appropriately designed, algorithms can contribute to human well-being by positively influencing user behaviour and preventing negative outcomes for humans. Some fundamental dilemmas apply, such as exemplified by the moral machine experiment~\citep{awad2018moral}, which algorithms often approach from a utilitarian or a duty ethics angle. Preventing ethical issues is a key concern of approaches such as IEEE~7000 that cover ethical value requirements~\citep{spiekermann2021expect}. While discussing ethics in more detail is beyond the scope of this paper, we would like to emphasize its importance in the context of algorithm engineering and point out that our framework can help identify where ethics may come into play. 

The \textit{ontological perspective} shows that ethical concerns are by no means limited to the design of algorithms. As the starting point is a real-world problem, one can already ask whether this real-world problem and the corresponding tasks are ethical. One prominent and widely discussed example is the task of creating so-called deepfakes, i.e., face-swapping technologies that enable the creation of fake images or videos that appear to be real~\cite{meskys2020regulating}. While there are several real-world problems that can ethically justify the deepfake creation task (e.g., bringing back dead actors), others are highly problematic (e.g., fake speeches of politicians or deep fake pornography). 

The \textit{epistemological perspective} points to the issue of creating knowledge about the design that provides insights into its ethical implications. Strictly speaking, such knowledge can be interpreted as a specific category of performance knowledge and to which extent corresponding requirements are satisfied. A difficulty of such an assessment is that dilemmas apply and ethical norms differ between individuals and cultural contexts~\citep{awad2018moral}.

The \textit{methodological perspective} particularly emphasizes the role of evaluation data. It has been widely discussed that data, especially when used for training of AI models, can lead to different types of biases and discrimination~\cite{ntoutsi2020bias}. Our framework further points at the relationship between evaluation data and the empirical knowledge we can generate about a design: data that is biased or not representative will lead to biased and non-representative knowledge. An aspect that might be less obvious is that the generation of evaluation data might be associated with ethical concerns as well. \citet{kshetri2021data} discusses the issues associated with large data  being manually labeled via crowdsourcing microtasks, often performed by individuals residing in developing countries who do not receive fair payment.        

We recognize that the aspects discussed above are a few out of many ethical concerns. We, however, believe that our framework can provide a useful angle to structure a disucssion on ethics in algorithm engineering.

\section{Conclusion}
\label{sec:conclusion}
In this paper, we presented a research framework for algorithm engineering. 
Our framework builds on three areas discussed in the philosophy of science: ontology, epistemology and methodology. In essence, \emph{ontology} 
describes algorithm engineering as being concerned with algorithmic problems, algorithmic tasks, algorithm designs and algorithm implementations. \emph{Epistemology} describes the body of know\-ledge of algorithm engineering as a collection of prescriptive and descriptive know\-ledge, residing in World 3 of Popper's Three Worlds model. \emph{Methodology} refers to the steps how we can systematically enhance our know\-ledge of specific algorithms. In this context, we identified nine validity concerns and discussed how researchers can respond to falsification.

Our framework has important methodological implications for researching algorithms in various areas of computer science. First, it offers a theoretical foundation for algorithm engineering grounded in the philosophy of science. This foundation clarifies how algorithms can be studied and which know\-ledge can be obtained. Second, our framework provides guidance for the methodological evaluation of newly proposed algorithms. Authors can use considerations that are related to the different validity concerns as a starting point. Third, this methodological grounding offers a framework for conducting research on algorithms in various areas of computer science in a way that supports an incremental research path. The concepts tied together in our framework might eventually help to foster know\-ledge transfer across sub-disciplines of computer science on how to research algorithms.

\section*{Acknowledgement}
We thank our numerous colleagues who commented on earlier versions of this manuscript.

\bibliographystyle{ACM-Reference-Format}

\end{document}